\documentclass[%
reprint,
superscriptaddress,
pra,
floatfix,
]{revtex4-2}

\usepackage{graphicx}
\usepackage{siunitx}
\usepackage{physics}
\usepackage[colorlinks=true,linkcolor=blue, citecolor=blue, urlcolor=blue]{hyperref}
\usepackage{comment}

\begin{document}
\title[A Tunable Transition Metal Dichalcogenide Entangled Photon-Pair Source]{A Tunable Transition Metal Dichalcogenide Entangled Photon-Pair Source}

\author{Maximilian~A.~Weissflog}
\email{maximilian.weissflog@uni-jena.de}
\affiliation{Institute of Applied Physics, Abbe Center of Photonics, Friedrich Schiller University Jena, Albert-Einstein-Straße 15, Jena, 07745, Germany}
\affiliation{Max Planck School of Photonics, Hans-Knöll-Straße 1, Jena, 07745, Germany}

\author{Anna~Fedotova}
\affiliation{Institute of Applied Physics, Abbe Center of Photonics, Friedrich Schiller University Jena, Albert-Einstein-Straße 15, Jena, 07745, Germany}
\affiliation{Institute of Solid State Physics, Friedrich Schiller University Jena, Helmholtzweg 3, Jena, 07743, Germany}

\author{Yilin~Tang}
\affiliation{School of Engineering, College of Science and Computer Science, The Australian National University, Canberra, Australian Capital Territory, Australia}

\author{Elkin~A.~Santos}

\author{Benjamin~Laudert}

\author{Saniya~Shinde}

\author{Fatemeh~Abtahi}

\author{Mina~Afsharnia}

\author{Inmaculada~Pérez~Pérez}

\author{Sebastian~Ritter}
\affiliation{Institute of Applied Physics, Abbe Center of Photonics, Friedrich Schiller University Jena, Albert-Einstein-Straße 15, Jena, 07745, Germany}

\author{Hao~Qin}
\affiliation{School of Engineering, College of Science and Computer Science, The Australian National University, Canberra, Australian Capital Territory, Australia}

\author{Jiri~Janousek}
\affiliation{School of Engineering, College of Science and Computer Science, The Australian National University, Canberra, Australian Capital Territory, Australia}
\affiliation{Australian Research Council Centre of Excellence for Quantum Computation and Communication Technology, The Australian National University, Canberra, Australian Capital Territory, Australia}

\author{Sai~Shradha}
\affiliation{Institute of Applied Physics, Abbe Center of Photonics, Friedrich Schiller University Jena, Albert-Einstein-Straße 15, Jena, 07745, Germany}
\affiliation{Institute for Condensed Matter Physics, Technical University of Darmstadt, Hochschulstraße. 6-8, Darmstadt, 64289, Germany}

\author{Isabelle~Staude}
\affiliation{Institute of Applied Physics, Abbe Center of Photonics, Friedrich Schiller University Jena, Albert-Einstein-Straße 15, Jena, 07745, Germany}
\affiliation{Institute of Solid State Physics, Friedrich Schiller University Jena, Helmholtzweg 3, Jena, 07743, Germany}

\author{Sina~Saravi}
\affiliation{Institute of Applied Physics, Abbe Center of Photonics, Friedrich Schiller University Jena, Albert-Einstein-Straße 15, Jena, 07745, Germany}

\author{Thomas~Pertsch}
\affiliation{Institute of Applied Physics, Abbe Center of Photonics, Friedrich Schiller University Jena, Albert-Einstein-Straße 15, Jena, 07745, Germany}
\affiliation{Fraunhofer Institute for Applied Optics and Precision Engineering IOF, Albert-Einstein-Straße 7, Jena, 07745, Germany}

\author{Frank~Setzpfandt}
\affiliation{Institute of Applied Physics, Abbe Center of Photonics, Friedrich Schiller University Jena, Albert-Einstein-Straße 15, Jena, 07745, Germany}
\affiliation{Fraunhofer Institute for Applied Optics and Precision Engineering IOF, Albert-Einstein-Straße 7, Jena, 07745, Germany}

\author{Yuerui~Lu}
\email{yuerui.lu@anu.edu.au}
\affiliation{School of Engineering, College of Science and Computer Science, The Australian National University, Canberra, Australian Capital Territory, Australia}
\affiliation{Australian Research Council Centre of Excellence for Quantum Computation and Communication Technology, The Australian National University, Canberra, Australian Capital Territory, Australia}

\author{Falk~Eilenberger}
\email{falk.eilenberger@uni-jena.de}
\affiliation{Institute of Applied Physics, Abbe Center of Photonics, Friedrich Schiller University Jena, Albert-Einstein-Straße 15, Jena, 07745, Germany}
\affiliation{Fraunhofer Institute for Applied Optics and Precision Engineering IOF, Albert-Einstein-Straße 7, Jena, 07745, Germany}

\begin{abstract}
	Entangled photon-pair sources are at the core of quantum applications like quantum key distribution, sensing, and imaging. Operation in space-limited and adverse environments such as in satellite-based and mobile communication requires robust entanglement sources with minimal size and weight requirements. Here, we meet this challenge by realizing a cubic micrometer scale entangled photon-pair source in a 3R-stacked transition metal dichalcogenide crystal. Its crystal symmetry enables the generation of polarization-entangled Bell states without additional components and provides tunability by simple control of the pump polarization. Remarkably, generation rate and state tuning are decoupled, leading to equal generation efficiency and no loss of entanglement. Combining transition metal dichalcogenides with monolithic cavities and integrated photonic circuitry or using quasi-phasematching opens the gate towards ultrasmall and scalable quantum devices.
\end{abstract}

\keywords{SPDC, TMD, Bell States, Entanglement, 3R-MoS\textsubscript{2}}

\maketitle

\section{Introduction}\label{sec1}

Entangled photon-pairs are the key-enabler for real-world implementations of quantum technologies like secure quantum key distribution \cite{scaraniSecurityPracticalQuantum2009,loSecureQuantumKey2014}, quantum sensing and imaging \cite{Basset19} as well as distributed quantum computing schemes \cite{fitzsimonsPrivateQuantumComputation2017}. Consequently, a large variety of entangled photon-pair sources (EPS) has been developed, often relying on spontaneous parametric down-conversion (SPDC) in second-order nonlinear crystals \cite{anwarEntangledPhotonpairSources2021}. Setting out from first EPS implementations based on single bulk crystals \cite{kwiatNewHighIntensitySource1995}, ever more complex source designs were developed to meet requirements for the degree of entanglement, quantum state fidelity, tunability, and brightness of the sources. Solutions to create entangled photon pairs are typically based on the interference of two distinct SPDC processes and range from using two crossed nonlinear crystals \cite{kwiatUltrabrightSourcePolarizationentangled1999} via combination of different down-conversion paths in Sagnac and linear interferometers \cite{anwarEntangledPhotonpairSources2021} to integrated photonic systems \cite{luChipintegratedVisibleTelecom2019}. Achieving a high degree of entanglement in such sources imposes very narrow tolerances on the properties of the different SPDC processes to allow the necessary coherent superposition. This technological challenge is further increased by the demand to operate these complex sources in adverse and space limited environments, such as a satellite \cite{yinSatellitebasedEntanglementDistribution2017,anwarDevelopmentCompactEntangled2022} or in customer-level applications like mobile-phone QKD, which also need a simple and scalable approach. Tunability between different entangled states is desirable for active quantum networks \cite{williamsQuantumSecretSharing2019}. In the light of these demands, a requirement list for an ideal EPS design would be the generation of high-fidelity, maximally entangled (Bell) states, switching between different entangled states, wide frequency coverage, and high brightness, combined with a robust, scalable design and small footprint using as few optical components as possible.

\begin{figure*}[t!]
	\centering
	\includegraphics[width=1\textwidth]{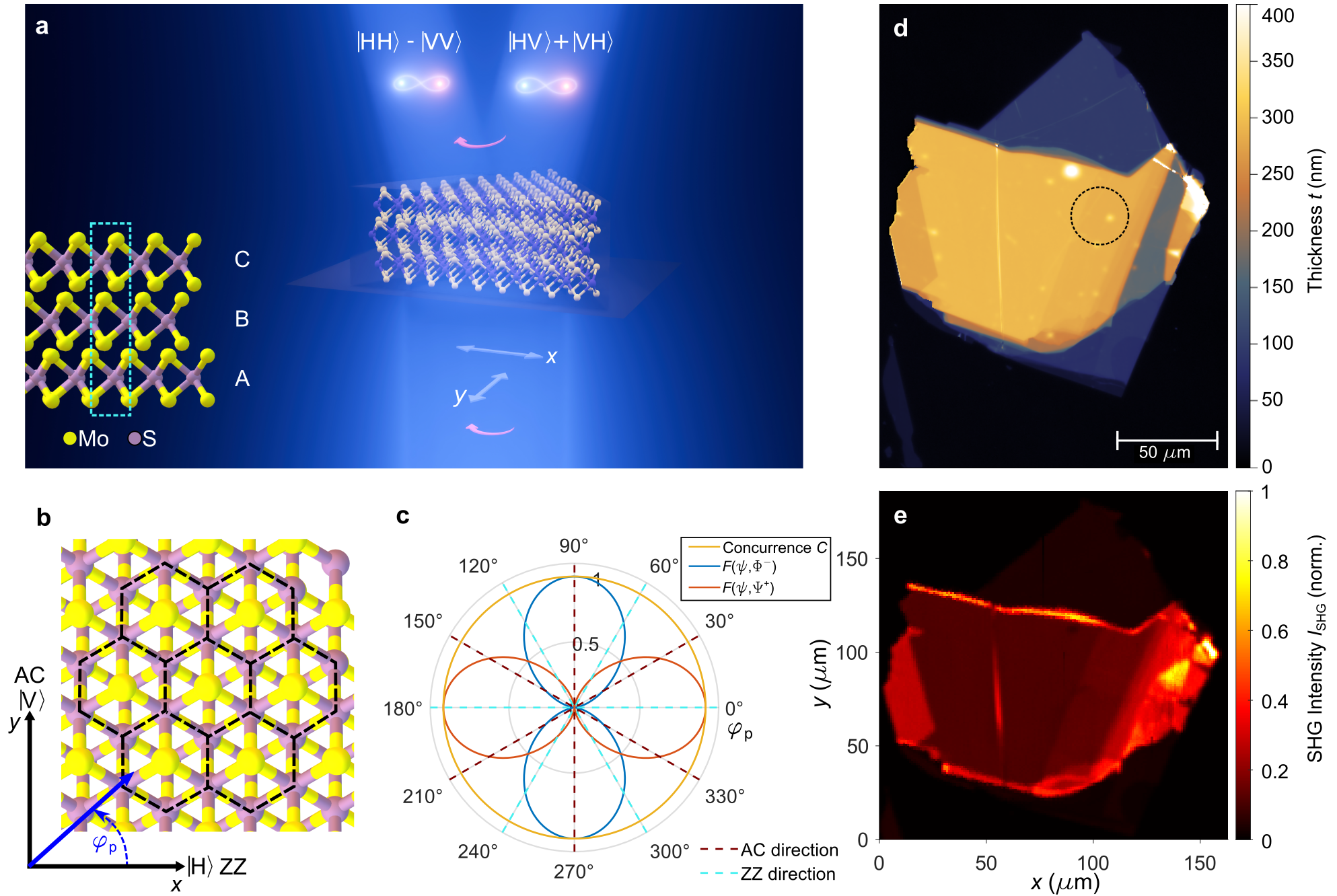}
	\caption{\textbf{Principle of entangled photon-pair generation in 3R-MoS\textsubscript{2}.} \textbf{a} A multilayer 3R-stack of MoS\textsubscript{2} generates pairs of polarization-entangled signal and idler photons via spontaneous parametric down-conversion (SPDC). Depending on the orientation of the pump polarization, different maximally entangled Bell states are generated. Lower inset: sketch of the ABC-stacking scheme in 3R-MoS\textsubscript{2}. \textbf{b} Top view of the crystalline structure of 3R-MoS\textsubscript{2} stack shown in (a). The unit cells of the crystal are highlighted with black, dashed lines. The $x$- and $y$- coordinates as well as the $\ket{\mathrm{H}}$- and $\ket{\mathrm{V}}$-polarization directions are aligned with the zigzag- and armchair directions of the crystal, respectively. The definition of the pump polarization angle $\varphi_{\mathrm{p}}$ is marked in blue. \textbf{c} Theoretical evolution of the fidelities $F(\psi,\Phi^{-})$ and $F(\psi,\Psi^{+})$ of the polarization state $\ket{\psi}$ with the Bell states $\ket{\Phi^{-}}=1/\sqrt{2}\left(\ket{\textrm{H}\textrm{H}}-\ket{\textrm{V}\textrm{V}}\right)$ and $\ket{\Psi^{+}}=1/\sqrt{2}\left(\ket{\textrm{H}\textrm{V}}+\ket{\textrm{V}\textrm{H}}\right)$, respectively, as well as the concurrence $C$ for a full rotation of the pump angle $\varphi_{\mathrm{p}}$. \textbf{d} Height-map of the investigated 3R-MoS\textsubscript{2} crystal obtained with a vertical scanning interferometer. The black circle marks the measurement region. \textbf{e} Map of second-harmonic generation (SHG) intensity for the 3R-MoS\textsubscript{2} crystal shown in (d), obtained for excitation at $2\times\SI{788}{nm}=\SI{1576}{nm}$.} 
	\label{fig:fig1}
\end{figure*}

In this work we demonstrate the core component of such a source, a submicron transition metal dichalcogenide (TMD) crystal that generates maximally entangled photon-pairs. To the best of our knowledge, this is the first realization of photon-pair generation in this material system. We use 3R-phase MoS\textsubscript{2} multilayer stacks with bulk-noncentrosymmetry \cite{shi3RMoS2Broken2017,xuCompactPhasematchedWaveguided2022,wagonerSecondharmonicGenerationMolybdenum1998}, which drastically increases the signal yield of nonlinear conversion \cite{malardObservationIntenseSecond2013,liProbingSymmetryProperties2013,klimmerAllopticalPolarizationAmplitude2021,trovatelloOpticalParametricAmplification2021,hsuSecondHarmonicGeneration2014} and simultaneously suppresses photoluminescence as compared to monolayer (ML) TMDs \cite{splendianiEmergingPhotoluminescenceMonolayer2010}. This is a decisive advantage compared to previous, inconclusive attempts to photon-pair generation in ML-TMDs
\cite{mariniConstraintsDownconversionAtomically2018,dinparastisalehSpontaneousParametricConversion2018}.
\newpage
Our photon-pair source based on 3R-phase molybdenum disulfide (MoS\textsubscript{2}) leverages the crystal symmetry of this van-der-Waals material to intrinsically create polarization entanglement. We demonstrate the broadband generation of maximally polarization entangled Bell states with a measured fidelity up to $\SI{96}{\percent}$. The need for external optical elements to create entanglement is obliterated, allowing to keep the optical system as simple as possible. Remarkably, the output quantum state of the TMD crystal can be easily tuned to different Bell and other maximally entangled states, all with the same generation efficiency. This property fundamentally stems from the crystal symmetry and goes beyond other recent demonstrations of thin-film nonlinear sources \cite{okothMicroscaleGenerationEntangled2019, sultanovFlatopticsGenerationBroadband2022,guoUltrathinQuantumLight2023}. The pair-generation rate of 3R-MoS$_2$ sources can be scaled to the required level e.g. through quasi-phasematching. Similar to periodic poling in ferroelectric materials, the nonlinearity in stacks of several multilayer 3R-MoS$_2$ crystals can be periodically modulated by suitably rotating consecutive crystals \cite{hsuSecondHarmonicGeneration2014,xuCompactPhasematchedWaveguided2022}. For specific technological applications requiring a high brightness of photon-pairs in defined spectral bands \cite{anwarDevelopmentCompactEntangled2022}, orders of magnitudes enhancement of the pair-rate in the desired range may be achieved by later on integrating the nonlinear TMD crystal into singly- or doubly resonant, monolithic cavities \cite{jeronimo-morenoTheoryCavityenhancedSpontaneous2010,slatteryBackgroundReviewCavityEnhanced2019}, an available technological process \cite{knopfIntegrationAtomicallyThin2019a,shradhaDoubleResonantCavity}. Based on our work, these readily developed technologies can in the future be combined to realize highly compact, flexible and robust entangled photon-pair sources based on TMDs.

\section{Results}\label{sec2}

\subsection{Fundamentals of Photon-Pair Generation and Polarization Entanglement in Transition Metal Dichalcogenides}\label{sec:pair_generation}

In the monolayer limit, TMDs with the structural form $MX_2$ ($M=$ \textit{Mo,W}; $X=$ \textit{S,Se}) show a strong second-order nonlinear response. Their crystal lattice is three-fold rotationally symmetric around the $z$-axis, corresponding to the point group $D_{3h}$. This leads to a $\hat{\chi}^{(2)}$ nonlinear tensor with non-vanishing elements $\chi^{(2)}_{\alpha\beta\gamma}=\chi^{(2)}_{yyy}=-\chi^{(2)}_{yxx}=-\chi^{(2)}_{xxy}=-\chi^{(2)}_{xyx}$ \cite{boydNonlinearOptics2008}. The $x$- and $y$-direction are defined based on the crystallographic zigzag (ZZ) and armchair (AC) directions, see Fig.~\ref{fig:fig1}(b). This nonlinear tensor couples electric fields with signal and idler frequencies $\omega_\mathrm{s},\omega_\mathrm{i}$ and polarization indices $\alpha,\beta$ to a higher-frequency pump field with $\omega_\mathrm{p}=\omega_\mathrm{s}+\omega_\mathrm{i}$ and polarization index $\gamma$. This enables classical three-wave mixing processes like second-harmonic generation (SHG) and sum-frequency generation (SFG) in TMDs, which were extensively studied \cite{malardObservationIntenseSecond2013,liProbingSymmetryProperties2013, klimmerAllopticalPolarizationAmplitude2021}. The same nonlinearity enables also SPDC, where due to vacuum fluctuations pump photons with frequency $\omega_\mathrm{p}$ spontaneously split into pairs of signal and idler photons with frequencies $\omega_\mathrm{s}$ and $\omega_\mathrm{i}$.

So far SPDC, which is the reverse process of SFG, could not be observed in TMDs \cite{dinparastisalehSpontaneousParametricConversion2018,mariniConstraintsDownconversionAtomically2018}. Using TMDs for SPDC would be particularly interesting, since their nonlinear tensor ensures that the generated signal and idler photons are intrinsically polarization-entangled. To demonstrate this, let us first consider a $y$-polarized pump photon. In this case, two pathways for down-conversion exist simultaneously, namely $\ket{y}_{\mathrm{pump}}\xrightarrow{\chi^{(2)}_{yyy}}\ket{yy}$ and $\ket{y}_{\mathrm{pump}}\xrightarrow{\chi^{(2)}_{xxy}}\ket{xx}$. Since both processes are coherently driven by the same pump photon, the ensuing quantum state is a coherent superposition of the two conversion possibilities with equal magnitudes as $\chi^{(2)}_{yyy}=-\chi^{(2)}_{xxy}$. The resulting polarization quantum state is $\ket{\Phi^{-}}=1/\sqrt{2}\left(\ket{xx}-\ket{yy}\right)$. This is one of the Bell states, a maximally entangled quantum state with high importance in quantum information processing. Equivalently, an $x$-polarized excitation results in a coherent superposition of the two down-conversion paths $\ket{x}_{\mathrm{pump}}\xrightarrow{\chi^{(2)}_{xyx}}\ket{xy}$ and $\ket{x}_{\mathrm{pump}}\xrightarrow{\chi^{(2)}_{yxx}}\ket{yx}$. This generates the Bell state $\ket{\Psi^{+}}=1/\sqrt{2}\left(\ket{xy}+\ket{yx}\right)$, again maximally entangled. For a pump polarization rotated by the angle $\varphi_{\mathrm{p}}$ with respect to the $x$-axis, the generated state is a superposition of these two Bell states in the form 

\begin{equation}
	\ket{\psi} = \frac{\sin(\varphi_{\mathrm{p}})}{\sqrt{2}} \left( \ket{\mathrm{H}\mathrm{H}} - \ket{\mathrm{V}\mathrm{V}} \right) + \frac{\cos(\varphi_{\mathrm{p}})}{\sqrt{2}} \left(\ket{\mathrm{H}\mathrm{V}} + \ket{\mathrm{V}\mathrm{H}} \right),
	\label{eq:quantum_state_TMD}
\end{equation}

where we have used now the horizontal $\ket{\mathrm{H}}$ and vertical $\ket{\mathrm{V}}$ basis states in the far-field for the notation. These are co-aligned with the crystallographic $x$-axis (ZZ) and the $y$-axis (AC), respectively (see Fig.~\ref{fig:fig1}(b)). Based on this general form of the quantum state, it is straightforward to characterize the entanglement of states that lie in between the $\Psi^{+}$-state for $x$-polarized excitation ($\varphi_{\mathrm{p}}=\SI{0}{\degree}$, horizontal) and the $\Phi^{-}$-state for $y$-polarized excitation ($\varphi_{\mathrm{p}}=\SI{90}{\degree}$, vertical). As entanglement measure we employ the concurrence $C$, a quantity ranging between $C=0$ for separable and $C=1$ for fully entangled states \cite{woottersEntanglementFormationArbitrary1998}. In Fig.~\ref{fig:fig1}(c) we plot the fidelity of the general state Eq.~\eqref{eq:quantum_state_TMD} with the Bell states $\Psi^{+}$- and $\Phi^{-}$ as well as the concurrence $C$ for a full rotation of the pump polarization angle $\varphi_{\mathrm{p}}$. While the state fidelities for the two Bell states peak at $\varphi_\mathrm{p}=\SI{0}{\degree}$ and $\SI{90}{\degree}$, the concurrence is $C=1$ for all $\varphi_{\mathrm{p}}$. In fact, the output polarization state from the TMD for any pump angle is always maximally entangled. For a full derivation and the used definitions of concurrence and fidelity, refer to supplementary section S1. Furthermore, analogous to the case of classical frequency up-conversion \cite{liProbingSymmetryProperties2013}, due to their crystal symmetry the spontaneous down-conversion rate in TMDs is independent of the pump polarization. Therefore, TMDs generate fully entangled polarization states which are tunable with constant efficiency by means of pump polarization change.

A drawback of ML-TMDs is the low absolute signal yield in nonlinear conversion due to the very small interaction length with the medium \cite{xuCompactPhasematchedWaveguided2022}. More promising for the practical implementation of nonlinear devices based on TMDs is the use of moderately thicker crystals, with a stacking scheme that still preserves non-centrosymmetry. One such material is the 3R-polytype of TMDs like MoS\textsubscript{2} \cite{wagonerSecondharmonicGenerationMolybdenum1998,shi3RMoS2Broken2017,xuCompactPhasematchedWaveguided2022}, where the layer-stacks are arranged in an ABC-ABC scheme that has no inversion centre (one stacking period consists of three layers, compare inset of Fig.~\ref{fig:fig1}(a)) \cite{shi3RMoS2Broken2017}. Since 3R-MoS\textsubscript{2} maintains the 3-fold rotational crystal symmetry and the related in-plane nonlinear tensor elements, it belongs to the $C_{3v}$ point group, also the thicker 3R-crystal stacks are suited to generate polarization-entangled quantum states. The signal yield, however, is much higher than for a monolayer. The out-of-plane, $z$-polarized nonlinear tensor components of 3R-MoS$_2$ practically do not contribute to the generated quantum state, refer to supplementary section S2 for a detailed discussion.

\begin{figure*}[t!]
	\centering
	\includegraphics[width=\textwidth]{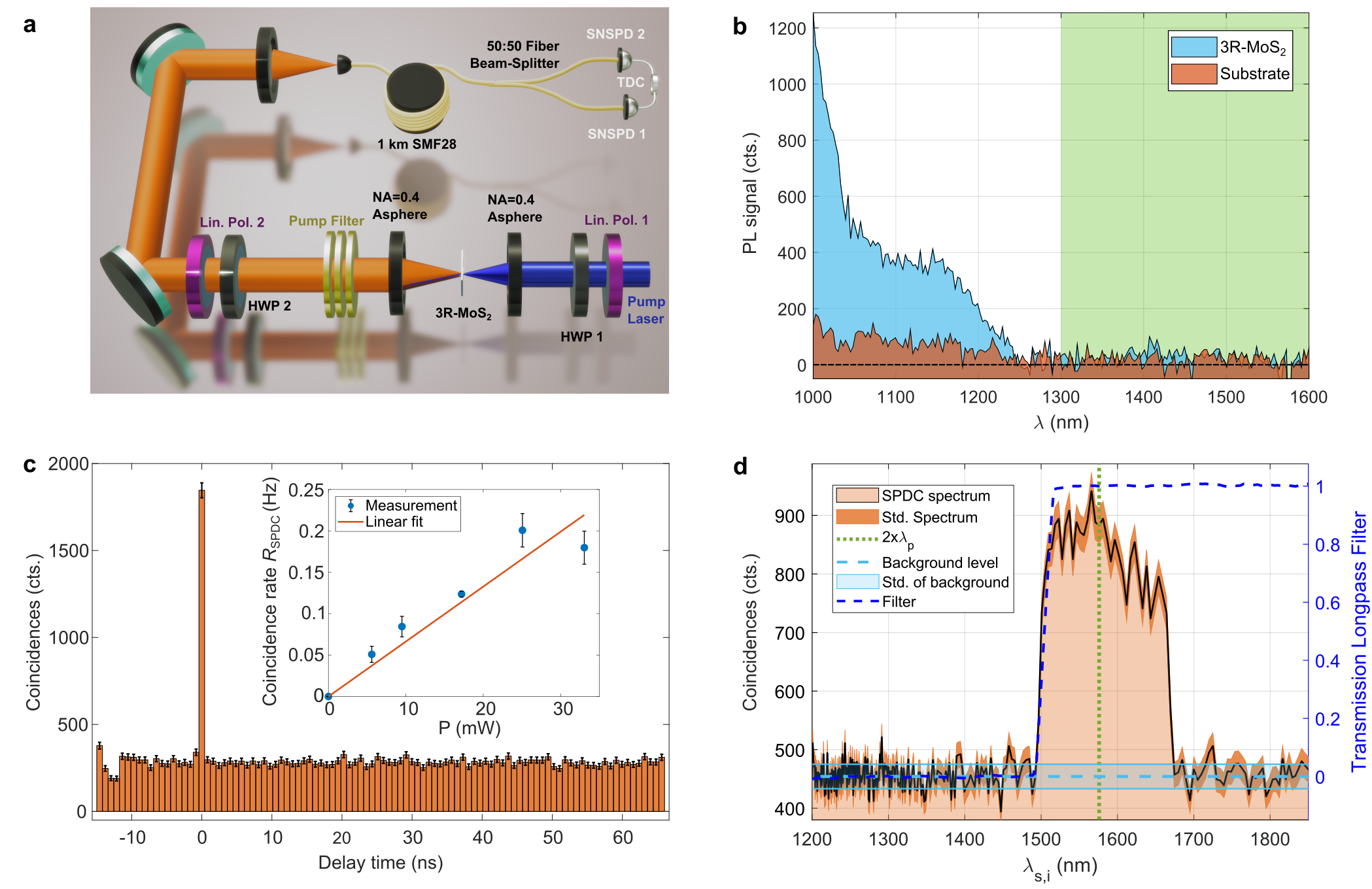}
	\caption{\textbf{Experimental observation of photon-pairs from 3R-MoS\textsubscript{2}.} \textbf{a} Hanbury Brown-Twiss interferometer for photon-pair correlation measurements: CW-diode laser at $\SI{788}{nm}$, linear polarizer (lin. pol.) and half-wave plate (HWP) for pump polarization preparation, aspheric lenses for focusing onto and collecting from the sample, long-pass interference filters (cut-on wavelength $\lambda=\SI{1100}{nm}$) for pump suppression, optional HWP and lin. pol. as polarization analyzer, single mode fiber (SMF) and 50:50 fiber beamsplitter, superconducting nanowire single-photon detector (SNSPD) and time-to-digital converter (TDC). \textbf{b} Photoluminescence (PL) spectrum from 3R-MoS\textsubscript{2} crystal and SiO\textsubscript{2} substrate excited at $\lambda_\mathrm{p}=\SI{788}{nm}$. \textbf{c} Coincidence histogram obtained from 3R-MoS\textsubscript{2} for excitation with $P=\SI{17.2}{mW}$ and integration time $\SI{3.5}{h}$, measured with a long-pass filter with cut-on wavelength $\SI{1500}{nm}$. The coincidence-to-accidental ratio (CAR) in this measurement is CAR$=5.5$. Error bars denote the statistical uncertainty. Inset: Measured SPDC coincidence rate (blue dots) for different pump powers and its linear fit (orange line). \textbf{d} Spectrum of SPDC photons measured using fiber spectroscopy. The dashed green line marks the degenerate SPDC wavelength $\lambda_{deg}=2\times\SI{788}{nm}=\SI{1576}{nm}$. The dark orange shaded areas mark the statistical uncertainty. The light blue lines show the level and standard deviation (std.) of the noise floor. The dashed, dark blue line is the transmission curve of the used long-pass filter that limits the SPDC spectrum.}
	\label{fig:fig2}
\end{figure*}

\newpage
\subsection{Experimental Photon-Pair Generation and Polarization Analysis}\label{sec:experimental_pair_generation}

Experimentally, we aim for photon-pair generation in the technically relevant telecom band in the near infrared around $\lambda_{\mathrm{s},\mathrm{i}}\approx\SI{1550}{nm}$. Using mechanical exfoliation, we fabricate a 3R-MoS\textsubscript{2} crystal with sub-wavelength thickness, see methods section. In Fig.~\ref{fig:fig1}(d) we show a height map of the crystal used as a photon-pair source in this work. For the SPDC measurement, we choose an area far away from the crystal edges and all cracks, which is important to minimize distortions of the nonlinear tensor induced by imperfections or strain \cite{shi3RMoS2Broken2017,mennelSecondHarmonicGeneration2019}. To further define the measurement area for the SPDC experiments, we first spatially map the SHG emitted by the crystal as shown in Fig.~\ref{fig:fig1}(e)). We choose the large area of $\SI{285}{\nm}$ height, see the marking in Fig.~\ref{fig:fig1}(d)), which shows a strong SHG signal in the center of the crystal. The signal yield from this crystal exceeds a ML-MoS\textsubscript{2} by more than three orders of magnitude (see supplementary section S3).

\begin{figure*}[t!]
	\centering
	\includegraphics[width=1\textwidth]{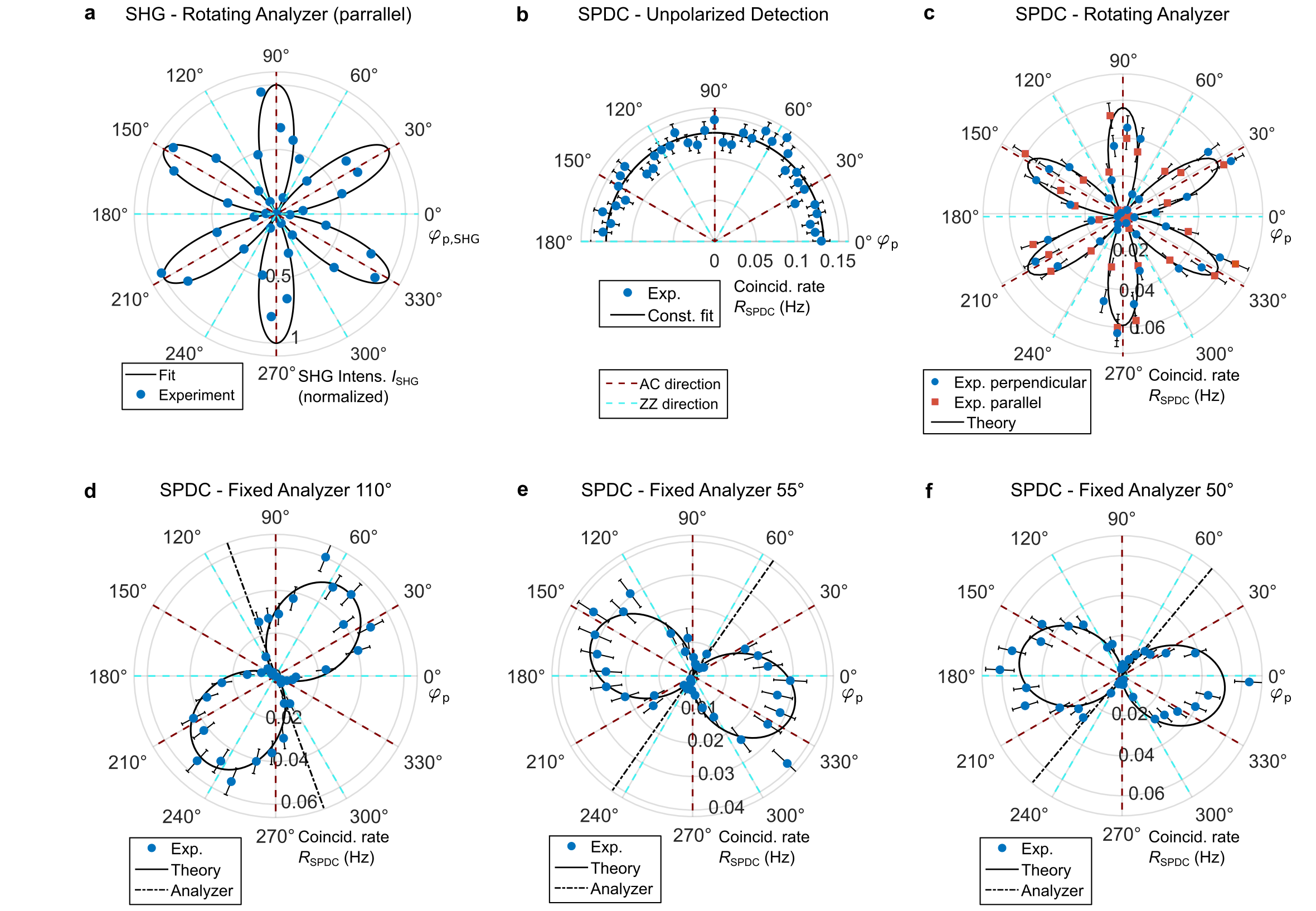}
	\caption{\textbf{Polarization analysis of photon-pairs.} \textbf{a} Polarization-resolved second-harmonic measurement (blue circles) for the rotation of the pump polarization angle $\varphi_{\mathrm{p,SHG}}$ and parallely polarized, co-rotating analyzer. The black curve is a fit with the theoretically expected dependence $I_\mathrm{SHG}\propto\sin^2(3\varphi_{\mathrm{p,SHG}})$. The armchair (AC), and zigzag (ZZ) directions are marked with dashed brown and cyan lines, respectively. \textbf{b} Measured photon-pair rate for rotation of the SPDC pump angle $\varphi_{\mathrm{p}}$ for unpolarized detection. The black curve marks a fit with the theoretically expected constant function. \textbf{c} Measured photon-pair rate for rotating pump polarization $\varphi_{\mathrm{p}}$ and detection through a co-rotating, perpendicularly oriented (blue circles) and parallely oriented (purple squares) analyzer. The in both cases expected $R_\mathrm{SPDC}\propto\sin^2(3\varphi_{\mathrm{p}})$ dependence (black curve) is overlayed. \textbf{d-f} Measured photon-pair rates for the rotating pump polarization $\varphi_{\mathrm{p}}$ and an analyzer fixed at (d) $\varphi_\mathrm{pol}=\SI{110}{\degree}$ ($\SI{20}{\degree}$ offset from AC-axis), (e) $\varphi_\mathrm{pol}=\SI{55}{\degree}$ ($\SI{25}{\degree}$ offset from AC-axis) and (f) $\varphi_\mathrm{pol}=\SI{50}{\degree}$ ($\SI{20}{\degree}$ offset from AC-axis). The dash-dotted black line marks the direction of the analyzer transmission axis, the solid dashed line is the theoretically expected dependence $R_\mathrm{SPDC}\propto\sin^2\left(2\varphi_\mathrm{pol}+\varphi_\mathrm{p}\right)$. Error bars show the statistical uncertainty.} 
	\label{fig:fig3}
\end{figure*}

For the photon-pair measurements we use an experimental setup with two fiber-coupled, time-correlated single-photon detectors as shown in Fig.~\ref{fig:fig2}(a). A pump beam with wavelength $\lambda_\mathrm{p}=\SI{788}{nm}$ is focused onto the air-exposed side of the 3R-MoS\textsubscript{2} sample, and photon-pairs are collected through the quartz substrate.

\newpage
In our correlation experiment, any other emission from the sample in the same wavelength region would potentially mask the entangled photon signal. In particular, strong photoluminescence as observed from direct bandgap transitions in ML-TMDs \cite{splendianiEmergingPhotoluminescenceMonolayer2010} could complicate the observation of photon-pairs \cite{dinparastisalehSpontaneousParametricConversion2018}. We measure photoluminescence from our sample under excitation at $\lambda_\mathrm{p}=\SI{788}{nm}$ from the same pump laser as in the SPDC experiments. We observe no photoluminescence signal distinguishable from the detector darkcounts beyond $\lambda = \SI{1300}{\nm}$ (see green shaded area Fig.~\ref{fig:fig2}(b)). This demonstrates the 3R-TMD's potential for low background photon-pair generation in the telecom wavelength band.

Consequently, we perform pair-correlation measurements and observe a pronounced coincidence peak, compare Fig.~\ref{fig:fig2}(c). After background subtraction, we measure $1563\pm43$ coincidence counts with a coincidence-to-accidental ratio (CAR) of CAR$=5.5\pm0.4$ for an integration time of $\SI{3.5}{h}$ and a pump power incident on the sample of $\SI{17.2}{mW}$. The maximum CAR we observe is CAR$=8.9\pm5.5$ for a pump power of $\SI{5.6}{mW}$ (see supplementary section S4.1). The value of CAR$>2$ together with the linear scaling of the coincidence rate with the pump power (see inset in Fig.~\ref{fig:fig2}(c)) is clear evidence for the SPDC origin of the coincidence peak. Furthermore, we measure the SPDC spectrum using fiber spectroscopy \cite{valenciaEntangledTwoPhotonWave2002a} (see methods).
\newpage
As expected for a non-phase matched, thin crystal, the SPDC spectrum is very broad \cite{okothMicroscaleGenerationEntangled2019,santiago-cruzEntangledPhotonsSubwavelength2021}, compare Fig.~\ref{fig:fig2}(d). In the experiment, the spectrum is limited only by the long-pass filter with cut-on $\SI{1500}{nm}$ used for suppression of residual photoluminescence (filter curve shown as dashed, dark blue line in Fig.~\ref{fig:fig2}(d)).

The specific form of the nonlinear tensor of 3R-MoS$_2$ leads to a characteristic dependence of the generated photon's polarization on the pump polarization, which we characterize next. As reference, we show in Fig.~\ref{fig:fig3}(a) a classical polarization-resolved second-harmonic measurement from 3R-MoS$_2$ observed through an analyzer, that is rotated in parallel to the pump polarization (see methods). The characteristic six-fold symmetric pattern is oriented along the AC crystal direction (dashed brown line in Fig.~\ref{fig:fig3}(a)) \cite{liProbingSymmetryProperties2013}.
For SPDC detection without a polarizer, we observe the expected constant coincidence rate, independent of the pump polarization (Fig. \ref{fig:fig3}(b)). We assign the small fluctuations in the measured rate mainly to the polarization sensitivity of our SNSPD detectors, which in the telecom range is significant \cite{changMultimodefibercoupledSuperconductingNanowire2019}. 

We then insert an analyzer in front of the fiber and simultaneously rotate the pump polarization angle $\varphi_{\mathrm{p}}$ and the analyzer angle $\varphi_{\mathrm{pol}}$ either in parallel (orange squares in Fig.~\ref{fig:fig3}(c)) or perpendicular configurations (blue dots in Fig.~\ref{fig:fig3}(c)). Both yield the characteristic six-fold pattern. Unlike in SHG however, they are co-aligned for both analyzer configurations \cite{liProbingSymmetryProperties2013}. We overlay this with the theoretically expected dependence for the SPDC rate $R_\mathrm{SPDC}\propto\sin^2\left(3\varphi_{\mathrm{p}}\right)$ and find a very good agreement with our measurements. Furthermore, by varying the pump polarization for several constant analyzer positions we obtain a two-lobed pattern (see Fig.~\ref{fig:fig3}(d-f)), again in good agreement with the theoretically expected dependence $R_\mathrm{SPDC}\propto\sin^2\left(2\varphi_\mathrm{pol}+\varphi_\mathrm{p}\right)$ (see supplementary section S1 for the derivation and S4.2 for raw coincidence histograms for all results in Fig.~\ref{fig:fig3}).

\subsection{Quantum-State Tomography and Bell-State Generation}\label{sec:quantum_state_tomography}

\begin{figure*}[t!]%
	\centering
	\includegraphics[width=1\textwidth]{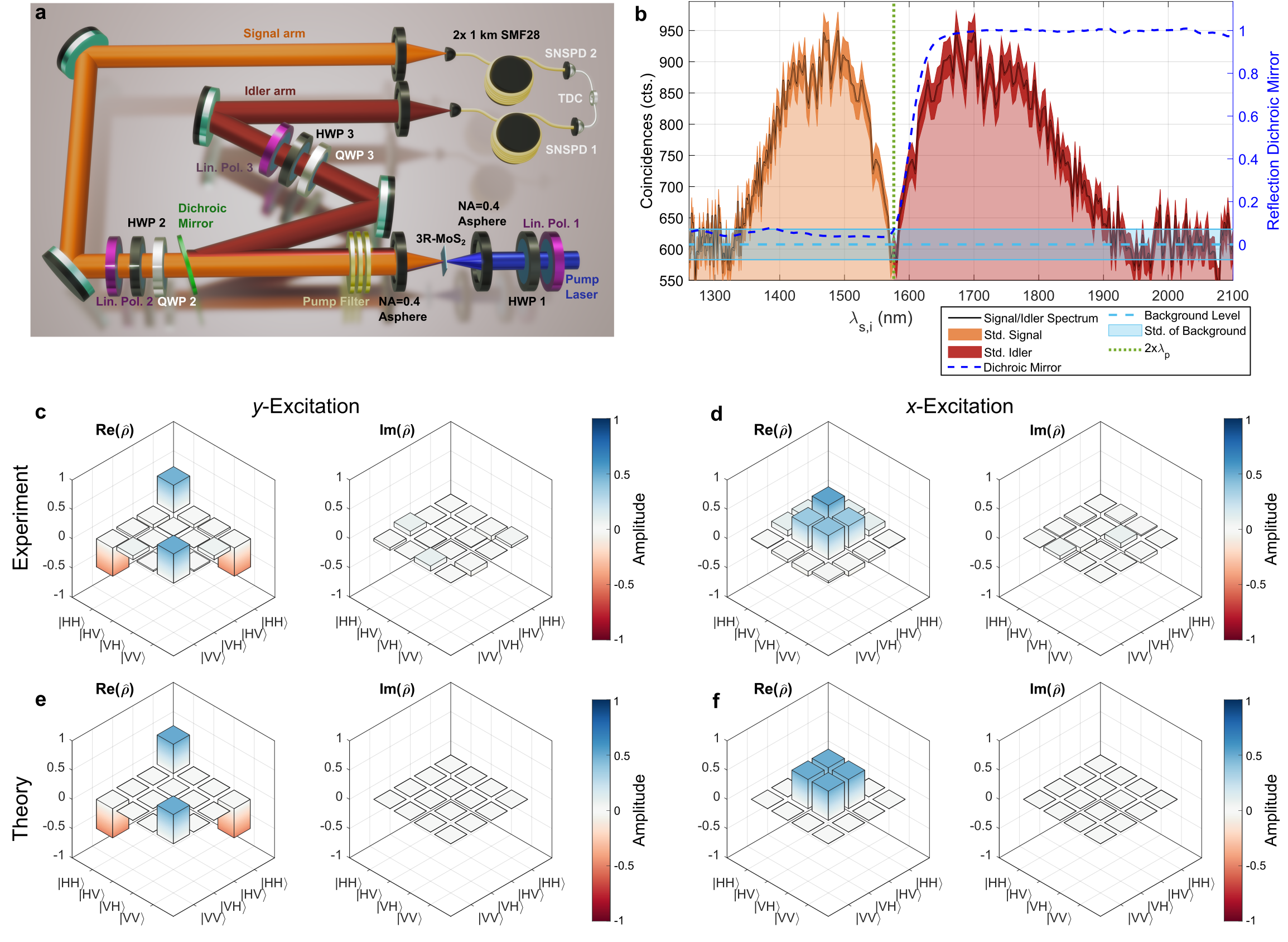}
	\caption{\textbf{Quantum-state tomography} \textbf{a} Experimental setup for quantum polarization-state tomography. A short-pass dichroic mirror (DM) with cut-on wavelength at $\lambda=\SI{1600}{nm}$ splits the signal (orange, upper arm) and idler (red, lower arm) frequency modes. Quantum-state tomography is performed with a set of quarter-wave plate (QWP), half-wave plate (HWP) and linear polarizer in each arm, then temporal correlations are measured as with the experiment described in Fig.~\ref{fig:fig2}(a). \textbf{b} Measured spectra of the signal (orange) and idler (red) frequency modes. The dip around the degenerate wavelength (dashed green line) is caused by the slight detuning of the cut-on wavelength of the dichroic mirror from the degenerate SPDC wavelength. The dark shaded areas mark the statistical uncertainty. The light blue lines show the level and standard deviation of the noise floor. The dashed, dark blue line is the reflection curve of the dichroic mirror. \textbf{c-d} Experimentally measured polarization density matrices $\hat{\rho}$ for (c) for $y$-, and (d) $x$-polarized pump. For both cases the real and imaginary parts $Re(\hat{\rho})$ and $Im(\hat{\rho})$, respectively, are shown. \textbf{e-f} Theoretically expected polarization density matrices obtained from fully vectorial Green's function calculations for $y$- and $x$-polarized excitations of 3R-MoS\textsubscript{2}, respectively.\\
	}\label{fig:fig4}
\end{figure*}

To completely characterize the generated polarization quantum state and to prove entanglement between signal and idler photons, we perform a tomographic measurement in two mutually unbiased polarization bases \cite{jamesMeasurementQubits2001}. To deterministically separate signal and idler photons, we insert a short-pass dichroic mirror with cut-on wavelength $\SI{1600}{nm}$ into the SPDC collection path as depicted in Fig.~\ref{fig:fig4}(a). The broadband signal and idler spectra and the beamsplitter reflection spectrum are shown in Fig.~\ref{fig:fig4}(b). Due to the collection in single-mode fibers and the use of an approximately non-polarizing dichroic mirror, the photon-pairs remain indistinguishable in all degrees of freedom but their frequency. Using a combination of waveplates and a linear polarizer in both paths allows to set two arbitrary, independent polarization bases. By performing projections into 16 different basis states, the density matrix $\hat{\rho}$ of the polarization quantum state is fully determined \cite{jamesMeasurementQubits2001} (see methods). We use an established maximum likelihood estimation method to determine a physically correct density matrix from measurements that are subject to noise and experimental uncertainties \cite{jamesMeasurementQubits2001,QuantumTomography2023}.

In Fig.~\ref{fig:fig4}(c-d) we show the real and imaginary parts of the experimentally obtained density matrices $\hat{\rho}$ for a $y$- (c) and $x$-polarized (d) pump (raw data in the supplementary section S4.3). Additionally we compute the theoretically expected state emitted from our 3R-MoS$_2$ with $t=\SI{285}{nm}$ based on fully vectorial Green's function calculations, taking into account the realistic conditions in our experiment for pump focusing, collection NA, collected SPDC bandwidth etc. (see methods). These calculations predict the generation of ideal Bell states, compare the theoretical density matrices in Fig.~\ref{fig:fig4}(e-f). This is closely matched by the experiment. For the $y$-polarized excitation, the measured density matrix has a fidelity of $F=0.96$ with a ${\ket{\Phi^{-}}=1/\sqrt{2}\left(\ket{\mathrm{H}\mathrm{H}}-\ket{\mathrm{V}\mathrm{V}}\right)}$ state and a concurrence of $C=0.967\pm0.002$, while for the $x$-polarized excitation the fidelity with a ${\ket{\Psi^{+}}=1/\sqrt{2}\left(\ket{\mathrm{H}\mathrm{V}}+\ket{\mathrm{V}\mathrm{H}}\right)}$ state is $F=0.84$ and the concurrence $C=0.87\pm0.02$.

\section{Discussion and Conclusion}\label{sec12}

In this work we observe for the first time SPDC in a transition metal dichalcogenide. We chose 3R-MoS\textsubscript{2} for our demonstration, because its strong nonlinearity is preserved in multi-layer stacks. Simultaneously, it is much less affected by photoluminescence than monolayer TMDs, which had prevented oberservation of SPDC in prior experiments. We demonstrate that TMDs intrinsically generate maximally entangled polarization Bell states.

Even more intriguingly, we show that for any linear pump polarization, a different, maximally entangled state is generated while the generation efficiency is independent of the pump polarization. This decoupling of entangled state tuning from the generation efficiency results in a highly flexible and easy to operate, tunable entangled photon-pair source. Since all these properties are directly derived from the crystal symmetry, no external optical components like interferometers etc. are needed for generating entanglement. This is the simplest conceivable, tunable entangled photon-pair source, a prerequisite for active quantum networks, which enable for instance multi-user quantum secret sharing \cite{williamsQuantumSecretSharing2019}. 

While we demonstrate here a prototype based on a single, thin 3R-MoS$_2$ crystal, the generation rate can be scaled to the required level, e.g. through quasi-phasematching. Similar to periodic poling of ferroelectric nonlinear materials \cite{anwarEntangledPhotonpairSources2021}, the nonlinearity of 3R-MoS$_2$ can be periodically poled by stacking several multilayer crystals with appropriate rotation angles between consecutive crystals \cite{xuCompactPhasematchedWaveguided2022,hsuSecondHarmonicGeneration2014}.
\newpage
With this, quasi-phasematching between pump, signal and idler wave is possible and the length of the 3R-TMD stack can be increased beyond one coherence length to match the photon-pair rate required in specific applications.

Another way of scaling up the source brightness is cavity integration. A cavity resonance at the pump wavelength effectively extends the interaction length with the nonlinear crystal, drastically enhancing the total pair-generation rate while resonances at the signal and idler wavelength strongly increase the spectral brightness in the desired frequency bands \cite{jeronimo-morenoTheoryCavityenhancedSpontaneous2010,slatteryBackgroundReviewCavityEnhanced2019}. The integration of TMDs into high-Q, monolithic cavities is a readily developed technology \cite{knopfIntegrationAtomicallyThin2019a} with doubly-resonant cavities in reach \cite{shradhaDoubleResonantCavity}. Also excitonic enhancement of the second-order nonlinear susceptibility is a promising avenue to further increase the source brightness \cite{xuanExcitonenhancedSpontaneousParametric2023}.

The demonstrated continuous tuning of the output state while maintaining maximal entanglement and a constant generation efficiency goes beyond what was shown with previously developed thin-film sources \cite{guoUltrathinQuantumLight2023,sultanovFlatopticsGenerationBroadband2022,santiago-cruzEntangledPhotonsSubwavelength2021}. Combined with the avenues for scaling the generation rate, this gives TMDs a clear advantage as nonlinear material platform for entangled photon-pair sources. Furthermore, the high refractive index of 3R-MoS$_2$ is well suited for strong field confinement when being nanostructured \cite{naumanTunableUnidirectionalNonlinear2021,zotevTransitionMetalDichalcogenide2022}, making it a perfect platform for hyper-entangled photon-pair generation in resonant nanostructures \cite{santiago-cruzPhotonPairsResonant2021,santiago-cruzResonantMetasurfacesGenerating2022,weissflog_nonlinear_2023}. Given that TMDs also withstand harsh conditions like found in space \cite{voglRadiationToleranceTwodimensional2019}, and can be easily integrated on top \cite{ngoInfibreSecondharmonicGeneration2022} or end-facet \cite{voglRoomTemperatureSingle2017a} of optical fibers \cite{ngoInfibreSecondharmonicGeneration2022}, waveguides \cite{kuppadakkathDirectGrowthMonolayer2022} and also metasurfaces \cite{lochnerHybridDielectricMetasurfaces2021}, we expect to see their immediate use in microscale or integrated photonic circuits and entangled photon-pair sources. Combined with their extremely low requirements for size and weight with highly scalable fabrication routes, they will enable quantum communication and quantum sensing for medical applications, life sciences, semiconductor industry, and consumer applications alike.

\section{Methods}

\subsection*{Sample Fabrication}\label{sec:sample_fabrication}

Bulk 3R-MoS\textsubscript{2} crystals were grown using the chemical-vapor transport technique. Subsequently, 3R-MoS\textsubscript{2} flakes were prepared on polydimethylsiloxane (PDMS), which begins with mechanical exfoliation of the crystals. Afterwards, the substrates were pre-treated in order to eliminate potential contamination, followed by a dry transfer method to transfer the 3R-MoS\textsubscript{2} flakes onto quartz substrates.

\subsection*{Thickness Characterization}\label{sec:thickness_characterization}

Sample thicknesses were characterized by a surface profiler and vertical scanning interferometry (VSI, Bruker Contour GT-K). The surface profiler and VSI are utilized to access the average thickness, surface roughness, and uniformity of the 3R-MoS\textsubscript{2} sample.

\subsection*{Polarization-Resolved SHG Measurements and SHG Mapping}\label{sec:SHG_mapping}

Polarization-resolved SHG measurements were carried out with the same setup as used for quantum measurements but working in reverse: the fundamental beam was incident from one of the collecting fibers and focused/collected with the same optics (see Fig.~\ref{fig:fig2}(a)). 
As a laser source, a tunable femtosecond laser (Coherent Chameleon with optical parametric oscillator Angewandte Physik und Elektronik GmbH APE OPO-X) with pulse width $\SI{100}{fs}$, repetition rate $\SI{80}{MHz}$, at a central wavelength $\SI{1576}{nm}$ and with FWHM $\SI{10}{nm}$ was used. The pump polarization was controlled with a half-wave plate, which rotated together with an analyzer placed in the collection path. Two long-pass filters installed in the collection path filtered out the fundamental beam, and SHG was detected with a sCMOS camera (Excelitas pco.edge 4.2 bi), all not shown in Fig.~\ref{fig:fig2}(a). The detected polarization of second harmonic was kept parallel to the pump polarization creating a characteristic six-fold pattern. This measurement was used as reference to identify the orientation of the AC and ZZ crystal directions in the 3R-MoS$_2$ sample.

SHG mapping was performed using a custom-built nonlinear microscopy setup. A fundamental beam from a tunable femtosecond laser (Spectra-Physics Mai Tai and optical parametric oscillator Inspire HF 100) with pulse width $\SI{100}{fs}$, repetition rate $\SI{80}{MHz}$, at a central wavelength $\SI{1576}{nm}$ and with FWHM $\SI{10}{nm}$ was focused onto the sample via a 20x with NA$= 0.4$ objective (Mitutoyo). The polarization of the fundamental beam was fixed to be parallel to the AC-axis of 3R-MoS\textsubscript{2}. The beam diameter reached $<\SI{6}{\micro\meter}$ FWHM. The SHG signal was collected via a 100x NA$=0.85$ objective (Zeiss) and passed through two short-pass filters to remove the fundamental beam. The sample was then scanned with $\SI{1}{\micro\meter}$ step-width on a motorized XYZ-stage (Newport), while the second-harmonic signal was detected using an EMCCD camera (iXon3, Andor).

In both experiments the excitation wavelength was chosen to correspond to the degenerate wavelength of SPDC pumped at $\lambda_\mathrm{p}=\SI{788}{nm}$.

\subsection*{Photon-Pair Correlation Measurements}\label{sec:correlation_measurements}

Photon-pair correlation measurements shown in Fig.~\ref{fig:fig2}~and~\ref{fig:fig3} were performed using the home-built Hanbury Brown-Twiss interferometer outlined in Fig.~\ref{fig:fig2}(a). Excitation photons from a continuous-wave laser at $\lambda_\mathrm{p}=\SI{788}{nm}$ (diode laser, Thorlabs) were sent through a linear polarizer and a half-wave plate for pump polarization control and focused onto the sample by an aspheric lens with numerical aperture NA$=0.4$ (Thorlabs). Subsequently, photon pairs were collected in transmission geometry using an identical lens. Pump photons were removed using three interference long-pass filters with cut-on wavelength $\SI{1100}{nm}$ (Thorlabs). For measurements shown in Fig.~\ref{fig:fig2} and\ref{fig:fig3} we also used a long-pass filter with cut-on wavelength $\SI{1500}{nm}$ to suppress any residual photoluminescence and to limit the photon-pair bandwidth to the operation range of the fiber beamsplitter. The photon-pairs were then coupled to single mode fibers (SMF28, Corning), separated using a broadband fiber beamsplitter with central wavelength $\SI{1550}{nm}$ (Thorlabs) and directed to two superconducting single photon detectors (SNSPD, Single Quantum Eos). Coincident detection events are registered with a time-correlator (qutools quTAG or ID Quantique ID800). For the polarization measurements with common analyzer for both photons in Fig.~\ref{fig:fig3}, we implement a rotating analyzer using an achromatic half-wave plate (Thorlabs) followed by a fixed linear polarizer, such that the polarization state in the detector fiber is always the same.
\newpage
This rules out the polarization dependence of the detectors. The total photon-pair detection efficiency of the setup $\eta_\mathrm{tot}$ follows from $\eta_\mathrm{tot}=T^2_\mathrm{opt}\times T^2_\mathrm{coupl}\times\eta_\mathrm{BS}\times\eta^2_\mathrm{detec}\times\eta_\mathrm{LP}\approx\SI{0.6}{\percent}$. For our setup, we estimate the following values: single photon optical transmission including lenses, filters, mirrors etc. $T_\mathrm{opt}\approx0.78$; single mode fiber coupling efficiency $\eta_\mathrm{coupl}\approx0.35$; fiber beam-splitter non-uniformity and probabilistic splitting $\eta_\mathrm{BS}\approx0.95^2\times0.5=0.45$; detection efficiency of SNSPDs at degenerate SPDC wavelength and averaged over different polarizations $\eta_\mathrm{detec}\approx0.6$; spectral detection factor for measurement with longpass filter $\SI{1500}{nm}$ $\eta_\mathrm{LP}=0.5$. The spectral detection factor takes into account that effectively half of the SPDC spectrum is detected when the longpass filter at $\SI{1500}{nm}$ is inserted (compare to spectrum in Fig.~\ref{fig:fig4}(b)).

\subsection*{Fiber Spectroscopy}

Fiber spectroscopy was carried out to measure the photon-pair spectrum by mapping the spectral information onto the temporal domain using a dispersive medium. In this work, the dispersive medium consisted of two spools of SMF-28 fiber (Corning), each with a length of $\SI{1}{km}$.

The fiber spectroscopy experiment was conducted in two distinct configurations. In the first scenario as shown in Fig.~\ref{fig:fig2}(a), the photon-pairs traveled through the same fiber spool. Following this, they were split using a 50:50 fiber beamsplitter before being detected by SNSPDs (Single Quantum Eos with timing jitter $\leq\SI{25}{ps}$). The arrival time differences were measured by a correlation electronics (qutools quTAG with timing jitter $\leq\SI{10}{ps}$). In the second configuration as shown in Fig.~\ref{fig:fig4}(a), the photon-pairs were initially separated via a dichroic mirror. Subsequently, a $\SI{1}{km}$ dispersive fiber spool was introduced into each of the photon pathways, before detection through the SNSPDs. The group-velocity dispersion of the fiber leads to a time delay between signal and idler photons, which can be mapped to their wavelength difference \cite{valenciaEntangledTwoPhotonWave2002a}.

\subsection*{Quantum-State Tomography}

For the tomographic measurement of the two-photon polarization quantum state, both photons have to be projected into mutually unbiased bases. For this, we first separated signal and idler photons based on their frequency in our Hanbury Brown-Twiss interferometer, see Fig.~\ref{fig:fig4}(a). We implemented a dichroic mirror by using the reflection of a slightly tilted short-pass interference filter (Edmund Optics) with cut-off wavelength $\SI{1600}{nm}$. In each collection arm of the correlation setup, an arbitrary polarization basis could be set using a sequence of achromatic quarter-wave plate, half-wave plate and linear polarizer (Thorlabs). During all changes of the polarization basis, the orientation of the linear polarizer was kept constant in order to avoid effects from the polarization sensitivity of the detectors. For a full reconstruction, the state has to be measured in 16 different basis configurations. Please refer to the supplementary information S4.3 for details on the chosen projection bases. We evaluated the measurements using a maximum likelihood method \cite{jamesMeasurementQubits2001,QuantumTomography2023}. The uncertainty of the state concurrence $C$, derived from the experimentally measured density matrix, was determined using a Monte Carlo approach \cite{altepeterQubitQuantumState2004}.

\subsection*{Green's Function Method for Pair-Generation in Layered Materials}

Our theoretical formalism is based on the Green’s function (GF) quantization approach for description of pair generation \cite{poddubny_generation_2016}, where the coincidence detection probability at different spatial coordinates for a signal and idler photon generated by a nonlinear source through SPDC takes the form:
\begin{align}\label{eq:prob_SI}
	&p_{GF}(\mathbf{r}_\mathrm{s},\mathbf{r}_\mathrm{i})\propto\bigg| \sum_{\alpha,\beta,\gamma}\sum_{\sigma_\mathrm{s},\sigma_I}d^*_{\mathrm{s},\sigma_\mathrm{s}}d^*_{\mathrm{i},\sigma_\mathrm{i}}\times\,\int\dd \mathbf{r}\,\chi^{(2)}_{\alpha\beta\gamma}(\mathbf{r})\times\\
	&\,E_{\mathrm{p},\gamma}(\mathbf{r},\omega_\mathrm{s}+\omega_\mathrm{i})\,G_{\sigma_\mathrm{s}\alpha}(\mathbf{r}_\mathrm{s},\mathbf{r},\omega_\mathrm{s})\,G_{\sigma_\mathrm{i}\beta}(\mathbf{r}_\mathrm{i},\mathbf{r},\omega_\mathrm{i})\bigg|^2\,,
\end{align}
where $\alpha$, $\beta$, and $\gamma$ indices run over the $x$, $y$, and $z$ directions. Here, $d_{\sigma}$ are the components of detection vector $\mathbf{d}$, where $\sigma=x,y,z$.
$E_{\mathrm{p},\gamma}(\mathbf{r})$ are the vector components of the complex-valued monochromatic pump with frequency $\omega_\mathrm{p}=\omega_\mathrm{s}+\omega_\mathrm{i}$. $G_{ij}(\mathbf{r,r'},\omega)$ are the tensor components of the electric GF. 
Finally, $\chi^{(2)}_{\alpha\beta\gamma}(\mathbf{r})$ are the components of the second-order nonlinear tensor. 

Here, the GF describes all the linear properties of the system, and is incorporated to the quantum formalism to include nonlinear processes that involve the generation of entangled photons such as SPDC. Due of the generality of the GF method, this formalism can describe any thickness of the 3R-MoS$_2$ nonlinear crystal, 
ultra-thin or thick, and it can be used to describe near- and far-field radiation in the non-paraxial regime \cite{Santos}. Remarkably, this formalism allows to keep track of any polarization and directionality effects in the pair-generation process, which makes it useful in reconstruction of polarization states of entangled photons \cite{weissflog_nonlinear_2023}. For modelling the 3R-MoS$_2$ crystal, we use the refractive index data provided in \cite{xuCompactPhasematchedWaveguided2022} and the relative magnitude of the nonlinear tensor elements $d_{22}$ and $d_{31}$ from \cite{wagonerSecondharmonicGenerationMolybdenum1998}. For a detailed discussion of the influence of the different tensor elements on the generated quantum states refer to the supplementary information section S2.

\newpage

\section*{Acknowledgments}

This work was funded by the Deutsche Forschungsgemeinschaft (DFG, German Research Foundation) through the International Research Training Group (IRTG) 2675 “Meta-ACTIVE”, project number 437527638, the Collaborative Research Center (CRC) 1375 “NOA”, through “MEGAPHONE” project number 505897284 and through the Emmy Noether Program, project number STA 1426/2-1. The authors further acknowledge funding from the Bundesministerium für Bildung und Forschung (BMBF, German Federal Ministry of Education and Research) under the project identifiers 13N14877, 13XP5053A, 16KISQ087A; and by the State of Thuringia (Quantum Hub Th\"uringen, 2021 FGI 0043). Furthermore, the authors acknowledge funding support from ANU PhD student scholarship, Australian Research Council (grant no. DP220102219, LE200100032) and ARC Centre of Excellence in Quantum Computation and Communication Technology (project number CE170100012). Additionally, this project has received funding from the European Union’s Horizon 2020 research and innovation programme under the H2020-FETOPEN-2018-2020 grant agreement no. [899673] (Metafast).

\section*{Author Contributions}

M.A.W. conceived the ideas, coordinated the measurements and theoretical modeling, designed the SPDC and tomography experiments, did the analytical calculations, analyzed the data, created the figures and wrote the first draft of the manuscript under the supervision of F.E., Y.L., T.P, I.S. and F.S. Y.T. fabricated all samples. M.A.W., A.F. and S.Shi. did the SPDC and polarization resolved SHG experiments. M.A.W. did the tomography experiments. E.S. did the GF calculations under supervision of S.S. A.F. measured the SHG map and S.R. measured the PL spectra. M.A., I.P.P, A.F. and M.A.W designed and calibrated the correlation fiber spectrometer. Y.T., H.Q. and J.J. designed and carried out the experiment for thickness dependent SHG efficiency. B.L. did analytical calculations and contributed to writing the first manuscript draft. F.A. and S.Shr. did preliminary SHG and PL analysis for 3R-MoS\textsubscript{2} samples. F.E., Y.L., T.P, I.S., S.S. and F.S. acquired funding and provided experimental resources. F.E., F.S., S.S., Y.L., A.F. provided major revisions of the manuscript. All authors discussed the results and contributed to the manuscript.

\bibliography{bibliography}

\begin{thebibliography}{51}%
\makeatletter
\providecommand \@ifxundefined [1]{%
 \@ifx{#1\undefined}
}%
\providecommand \@ifnum [1]{%
 \ifnum #1\expandafter \@firstoftwo
 \else \expandafter \@secondoftwo
 \fi
}%
\providecommand \@ifx [1]{%
 \ifx #1\expandafter \@firstoftwo
 \else \expandafter \@secondoftwo
 \fi
}%
\providecommand \natexlab [1]{#1}%
\providecommand \enquote  [1]{``#1''}%
\providecommand \bibnamefont  [1]{#1}%
\providecommand \bibfnamefont [1]{#1}%
\providecommand \citenamefont [1]{#1}%
\providecommand \href@noop [0]{\@secondoftwo}%
\providecommand \href [0]{\begingroup \@sanitize@url \@href}%
\providecommand \@href[1]{\@@startlink{#1}\@@href}%
\providecommand \@@href[1]{\endgroup#1\@@endlink}%
\providecommand \@sanitize@url [0]{\catcode `\\12\catcode `\$12\catcode `\&12\catcode `\#12\catcode `\^12\catcode `\_12\catcode `\%12\relax}%
\providecommand \@@startlink[1]{}%
\providecommand \@@endlink[0]{}%
\providecommand \url  [0]{\begingroup\@sanitize@url \@url }%
\providecommand \@url [1]{\endgroup\@href {#1}{\urlprefix }}%
\providecommand \urlprefix  [0]{URL }%
\providecommand \Eprint [0]{\href }%
\providecommand \doibase [0]{https://doi.org/}%
\providecommand \selectlanguage [0]{\@gobble}%
\providecommand \bibinfo  [0]{\@secondoftwo}%
\providecommand \bibfield  [0]{\@secondoftwo}%
\providecommand \translation [1]{[#1]}%
\providecommand \BibitemOpen [0]{}%
\providecommand \bibitemStop [0]{}%
\providecommand \bibitemNoStop [0]{.\EOS\space}%
\providecommand \EOS [0]{\spacefactor3000\relax}%
\providecommand \BibitemShut  [1]{\csname bibitem#1\endcsname}%
\let\auto@bib@innerbib\@empty
\bibitem [{\citenamefont {Scarani}\ \emph {et~al.}(2009)\citenamefont {Scarani}, \citenamefont {Bechmann-Pasquinucci}, \citenamefont {Cerf}, \citenamefont {Dušek}, \citenamefont {Lütkenhaus},\ and\ \citenamefont {Peev}}]{scaraniSecurityPracticalQuantum2009}%
  \BibitemOpen
  \bibfield  {author} {\bibinfo {author} {\bibfnamefont {V.}~\bibnamefont {Scarani}}, \bibinfo {author} {\bibfnamefont {H.}~\bibnamefont {Bechmann-Pasquinucci}}, \bibinfo {author} {\bibfnamefont {N.~J.}\ \bibnamefont {Cerf}}, \bibinfo {author} {\bibfnamefont {M.}~\bibnamefont {Dušek}}, \bibinfo {author} {\bibfnamefont {N.}~\bibnamefont {Lütkenhaus}},\ and\ \bibinfo {author} {\bibfnamefont {M.}~\bibnamefont {Peev}},\ }\bibfield  {title} {\bibinfo {title} {The security of practical quantum key distribution},\ }\href {https://doi.org/10.1103/RevModPhys.81.1301} {\bibfield  {journal} {\bibinfo  {journal} {Reviews of Modern Physics}\ }\textbf {\bibinfo {volume} {81}},\ \bibinfo {pages} {1301} (\bibinfo {year} {2009})}\BibitemShut {NoStop}%
\bibitem [{\citenamefont {Lo}\ \emph {et~al.}(2014)\citenamefont {Lo}, \citenamefont {Curty},\ and\ \citenamefont {Tamaki}}]{loSecureQuantumKey2014}%
  \BibitemOpen
  \bibfield  {author} {\bibinfo {author} {\bibfnamefont {H.-K.}\ \bibnamefont {Lo}}, \bibinfo {author} {\bibfnamefont {M.}~\bibnamefont {Curty}},\ and\ \bibinfo {author} {\bibfnamefont {K.}~\bibnamefont {Tamaki}},\ }\bibfield  {title} {\bibinfo {title} {Secure quantum key distribution},\ }\href {https://doi.org/10.1038/nphoton.2014.149} {\bibfield  {journal} {\bibinfo  {journal} {Nature Photonics}\ }\textbf {\bibinfo {volume} {8}},\ \bibinfo {pages} {595} (\bibinfo {year} {2014})}\BibitemShut {NoStop}%
\bibitem [{\citenamefont {Gilaberte~Basset}\ \emph {et~al.}(2019)\citenamefont {Gilaberte~Basset}, \citenamefont {Setzpfandt}, \citenamefont {Steinlechner}, \citenamefont {Beckert}, \citenamefont {Pertsch},\ and\ \citenamefont {Gräfe}}]{Basset19}%
  \BibitemOpen
  \bibfield  {author} {\bibinfo {author} {\bibfnamefont {M.}~\bibnamefont {Gilaberte~Basset}}, \bibinfo {author} {\bibfnamefont {F.}~\bibnamefont {Setzpfandt}}, \bibinfo {author} {\bibfnamefont {F.}~\bibnamefont {Steinlechner}}, \bibinfo {author} {\bibfnamefont {E.}~\bibnamefont {Beckert}}, \bibinfo {author} {\bibfnamefont {T.}~\bibnamefont {Pertsch}},\ and\ \bibinfo {author} {\bibfnamefont {M.}~\bibnamefont {Gräfe}},\ }\bibfield  {title} {\bibinfo {title} {Perspectives for applications of quantum imaging},\ }\href {https://doi.org/10.1002/lpor.201900097} {\bibfield  {journal} {\bibinfo  {journal} {Laser \& Photonics Reviews}\ }\textbf {\bibinfo {volume} {13}},\ \bibinfo {pages} {1900097} (\bibinfo {year} {2019})}\BibitemShut {NoStop}%
\bibitem [{\citenamefont {Fitzsimons}(2017)}]{fitzsimonsPrivateQuantumComputation2017}%
  \BibitemOpen
  \bibfield  {author} {\bibinfo {author} {\bibfnamefont {J.~F.}\ \bibnamefont {Fitzsimons}},\ }\bibfield  {title} {\bibinfo {title} {Private quantum computation: An introduction to blind quantum computing and related protocols},\ }\href {https://doi.org/10.1038/s41534-017-0025-3} {\bibfield  {journal} {\bibinfo  {journal} {npj Quantum Information}\ }\textbf {\bibinfo {volume} {3}},\ \bibinfo {pages} {1} (\bibinfo {year} {2017})}\BibitemShut {NoStop}%
\bibitem [{\citenamefont {Anwar}\ \emph {et~al.}(2021)\citenamefont {Anwar}, \citenamefont {Perumangatt}, \citenamefont {Steinlechner}, \citenamefont {Jennewein},\ and\ \citenamefont {Ling}}]{anwarEntangledPhotonpairSources2021}%
  \BibitemOpen
  \bibfield  {author} {\bibinfo {author} {\bibfnamefont {A.}~\bibnamefont {Anwar}}, \bibinfo {author} {\bibfnamefont {C.}~\bibnamefont {Perumangatt}}, \bibinfo {author} {\bibfnamefont {F.}~\bibnamefont {Steinlechner}}, \bibinfo {author} {\bibfnamefont {T.}~\bibnamefont {Jennewein}},\ and\ \bibinfo {author} {\bibfnamefont {A.}~\bibnamefont {Ling}},\ }\bibfield  {title} {\bibinfo {title} {Entangled photon-pair sources based on three-wave mixing in bulk crystals},\ }\href {https://doi.org/10.1063/5.0023103} {\bibfield  {journal} {\bibinfo  {journal} {Review of Scientific Instruments}\ }\textbf {\bibinfo {volume} {92}},\ \bibinfo {pages} {041101} (\bibinfo {year} {2021})}\BibitemShut {NoStop}%
\bibitem [{\citenamefont {Kwiat}\ \emph {et~al.}(1995)\citenamefont {Kwiat}, \citenamefont {Mattle}, \citenamefont {Weinfurter}, \citenamefont {Zeilinger}, \citenamefont {Sergienko},\ and\ \citenamefont {Shih}}]{kwiatNewHighIntensitySource1995}%
  \BibitemOpen
  \bibfield  {author} {\bibinfo {author} {\bibfnamefont {P.~G.}\ \bibnamefont {Kwiat}}, \bibinfo {author} {\bibfnamefont {K.}~\bibnamefont {Mattle}}, \bibinfo {author} {\bibfnamefont {H.}~\bibnamefont {Weinfurter}}, \bibinfo {author} {\bibfnamefont {A.}~\bibnamefont {Zeilinger}}, \bibinfo {author} {\bibfnamefont {A.~V.}\ \bibnamefont {Sergienko}},\ and\ \bibinfo {author} {\bibfnamefont {Y.}~\bibnamefont {Shih}},\ }\bibfield  {title} {\bibinfo {title} {New {{High-Intensity Source}} of {{Polarization-Entangled Photon Pairs}}},\ }\href {https://doi.org/10.1103/PhysRevLett.75.4337} {\bibfield  {journal} {\bibinfo  {journal} {Physical Review Letters}\ }\textbf {\bibinfo {volume} {75}},\ \bibinfo {pages} {4337} (\bibinfo {year} {1995})}\BibitemShut {NoStop}%
\bibitem [{\citenamefont {Kwiat}\ \emph {et~al.}(1999)\citenamefont {Kwiat}, \citenamefont {Waks}, \citenamefont {White}, \citenamefont {Appelbaum},\ and\ \citenamefont {Eberhard}}]{kwiatUltrabrightSourcePolarizationentangled1999}%
  \BibitemOpen
  \bibfield  {author} {\bibinfo {author} {\bibfnamefont {P.~G.}\ \bibnamefont {Kwiat}}, \bibinfo {author} {\bibfnamefont {E.}~\bibnamefont {Waks}}, \bibinfo {author} {\bibfnamefont {A.~G.}\ \bibnamefont {White}}, \bibinfo {author} {\bibfnamefont {I.}~\bibnamefont {Appelbaum}},\ and\ \bibinfo {author} {\bibfnamefont {P.~H.}\ \bibnamefont {Eberhard}},\ }\bibfield  {title} {\bibinfo {title} {Ultrabright source of polarization-entangled photons},\ }\href {https://doi.org/10.1103/PhysRevA.60.R773} {\bibfield  {journal} {\bibinfo  {journal} {Physical Review A}\ }\textbf {\bibinfo {volume} {60}},\ \bibinfo {pages} {R773} (\bibinfo {year} {1999})}\BibitemShut {NoStop}%
\bibitem [{\citenamefont {Lu}\ \emph {et~al.}(2019)\citenamefont {Lu}, \citenamefont {Li}, \citenamefont {Westly}, \citenamefont {Moille}, \citenamefont {Singh}, \citenamefont {Anant},\ and\ \citenamefont {Srinivasan}}]{luChipintegratedVisibleTelecom2019}%
  \BibitemOpen
  \bibfield  {author} {\bibinfo {author} {\bibfnamefont {X.}~\bibnamefont {Lu}}, \bibinfo {author} {\bibfnamefont {Q.}~\bibnamefont {Li}}, \bibinfo {author} {\bibfnamefont {D.~A.}\ \bibnamefont {Westly}}, \bibinfo {author} {\bibfnamefont {G.}~\bibnamefont {Moille}}, \bibinfo {author} {\bibfnamefont {A.}~\bibnamefont {Singh}}, \bibinfo {author} {\bibfnamefont {V.}~\bibnamefont {Anant}},\ and\ \bibinfo {author} {\bibfnamefont {K.}~\bibnamefont {Srinivasan}},\ }\bibfield  {title} {\bibinfo {title} {Chip-integrated visible–telecom entangled photon pair source for quantum communication},\ }\href {https://doi.org/10.1038/s41567-018-0394-3} {\bibfield  {journal} {\bibinfo  {journal} {Nature Physics}\ }\textbf {\bibinfo {volume} {15}},\ \bibinfo {pages} {373} (\bibinfo {year} {2019})}\BibitemShut {NoStop}%
\bibitem [{\citenamefont {Yin}\ \emph {et~al.}(2017)\citenamefont {Yin}, \citenamefont {Cao}, \citenamefont {Li}, \citenamefont {Liao}, \citenamefont {Zhang}, \citenamefont {Ren}, \citenamefont {Cai}, \citenamefont {Liu}, \citenamefont {Li}, \citenamefont {Dai}, \citenamefont {Li}, \citenamefont {Lu}, \citenamefont {Gong}, \citenamefont {Xu}, \citenamefont {Li}, \citenamefont {Li}, \citenamefont {Yin}, \citenamefont {Jiang}, \citenamefont {Li}, \citenamefont {Jia}, \citenamefont {Ren}, \citenamefont {He}, \citenamefont {Zhou}, \citenamefont {Zhang}, \citenamefont {Wang}, \citenamefont {Chang}, \citenamefont {Zhu}, \citenamefont {Liu}, \citenamefont {Chen}, \citenamefont {Lu}, \citenamefont {Shu}, \citenamefont {Peng}, \citenamefont {Wang},\ and\ \citenamefont {Pan}}]{yinSatellitebasedEntanglementDistribution2017}%
  \BibitemOpen
  \bibfield  {author} {\bibinfo {author} {\bibfnamefont {J.}~\bibnamefont {Yin}}, \bibinfo {author} {\bibfnamefont {Y.}~\bibnamefont {Cao}}, \bibinfo {author} {\bibfnamefont {Y.-H.}\ \bibnamefont {Li}}, \bibinfo {author} {\bibfnamefont {S.-K.}\ \bibnamefont {Liao}}, \bibinfo {author} {\bibfnamefont {L.}~\bibnamefont {Zhang}}, \bibinfo {author} {\bibfnamefont {J.-G.}\ \bibnamefont {Ren}}, \bibinfo {author} {\bibfnamefont {W.-Q.}\ \bibnamefont {Cai}}, \bibinfo {author} {\bibfnamefont {W.-Y.}\ \bibnamefont {Liu}}, \bibinfo {author} {\bibfnamefont {B.}~\bibnamefont {Li}}, \bibinfo {author} {\bibfnamefont {H.}~\bibnamefont {Dai}}, \bibinfo {author} {\bibfnamefont {G.-B.}\ \bibnamefont {Li}}, \bibinfo {author} {\bibfnamefont {Q.-M.}\ \bibnamefont {Lu}}, \bibinfo {author} {\bibfnamefont {Y.-H.}\ \bibnamefont {Gong}}, \bibinfo {author} {\bibfnamefont {Y.}~\bibnamefont {Xu}}, \bibinfo {author} {\bibfnamefont {S.-L.}\ \bibnamefont {Li}}, \bibinfo {author} {\bibfnamefont {F.-Z.}\ \bibnamefont {Li}}, \bibinfo {author}
  {\bibfnamefont {Y.-Y.}\ \bibnamefont {Yin}}, \bibinfo {author} {\bibfnamefont {Z.-Q.}\ \bibnamefont {Jiang}}, \bibinfo {author} {\bibfnamefont {M.}~\bibnamefont {Li}}, \bibinfo {author} {\bibfnamefont {J.-J.}\ \bibnamefont {Jia}}, \bibinfo {author} {\bibfnamefont {G.}~\bibnamefont {Ren}}, \bibinfo {author} {\bibfnamefont {D.}~\bibnamefont {He}}, \bibinfo {author} {\bibfnamefont {Y.-L.}\ \bibnamefont {Zhou}}, \bibinfo {author} {\bibfnamefont {X.-X.}\ \bibnamefont {Zhang}}, \bibinfo {author} {\bibfnamefont {N.}~\bibnamefont {Wang}}, \bibinfo {author} {\bibfnamefont {X.}~\bibnamefont {Chang}}, \bibinfo {author} {\bibfnamefont {Z.-C.}\ \bibnamefont {Zhu}}, \bibinfo {author} {\bibfnamefont {N.-L.}\ \bibnamefont {Liu}}, \bibinfo {author} {\bibfnamefont {Y.-A.}\ \bibnamefont {Chen}}, \bibinfo {author} {\bibfnamefont {C.-Y.}\ \bibnamefont {Lu}}, \bibinfo {author} {\bibfnamefont {R.}~\bibnamefont {Shu}}, \bibinfo {author} {\bibfnamefont {C.-Z.}\ \bibnamefont {Peng}}, \bibinfo {author} {\bibfnamefont {J.-Y.}\
  \bibnamefont {Wang}},\ and\ \bibinfo {author} {\bibfnamefont {J.-W.}\ \bibnamefont {Pan}},\ }\bibfield  {title} {\bibinfo {title} {Satellite-based entanglement distribution over 1200 kilometers},\ }\href {https://doi.org/10.1126/science.aan3211} {\bibfield  {journal} {\bibinfo  {journal} {Science}\ }\textbf {\bibinfo {volume} {356}},\ \bibinfo {pages} {1140} (\bibinfo {year} {2017})}\BibitemShut {NoStop}%
\bibitem [{\citenamefont {Anwar}\ \emph {et~al.}(2022)\citenamefont {Anwar}, \citenamefont {Perumangatt}, \citenamefont {Villar}, \citenamefont {Lohrmann},\ and\ \citenamefont {Ling}}]{anwarDevelopmentCompactEntangled2022}%
  \BibitemOpen
  \bibfield  {author} {\bibinfo {author} {\bibfnamefont {A.}~\bibnamefont {Anwar}}, \bibinfo {author} {\bibfnamefont {C.}~\bibnamefont {Perumangatt}}, \bibinfo {author} {\bibfnamefont {A.}~\bibnamefont {Villar}}, \bibinfo {author} {\bibfnamefont {A.}~\bibnamefont {Lohrmann}},\ and\ \bibinfo {author} {\bibfnamefont {A.}~\bibnamefont {Ling}},\ }\bibfield  {title} {\bibinfo {title} {Development of compact entangled photon-pair sources for satellites},\ }\href {https://doi.org/10.1063/5.0109702} {\bibfield  {journal} {\bibinfo  {journal} {Applied Physics Letters}\ }\textbf {\bibinfo {volume} {121}},\ \bibinfo {pages} {220503} (\bibinfo {year} {2022})}\BibitemShut {NoStop}%
\bibitem [{\citenamefont {Williams}\ \emph {et~al.}(2019)\citenamefont {Williams}, \citenamefont {Lukens}, \citenamefont {Peters}, \citenamefont {Qi},\ and\ \citenamefont {Grice}}]{williamsQuantumSecretSharing2019}%
  \BibitemOpen
  \bibfield  {author} {\bibinfo {author} {\bibfnamefont {B.~P.}\ \bibnamefont {Williams}}, \bibinfo {author} {\bibfnamefont {J.~M.}\ \bibnamefont {Lukens}}, \bibinfo {author} {\bibfnamefont {N.~A.}\ \bibnamefont {Peters}}, \bibinfo {author} {\bibfnamefont {B.}~\bibnamefont {Qi}},\ and\ \bibinfo {author} {\bibfnamefont {W.~P.}\ \bibnamefont {Grice}},\ }\bibfield  {title} {\bibinfo {title} {Quantum secret sharing with polarization-entangled photon pairs},\ }\href {https://doi.org/10.1103/PhysRevA.99.062311} {\bibfield  {journal} {\bibinfo  {journal} {Physical Review A}\ }\textbf {\bibinfo {volume} {99}},\ \bibinfo {pages} {062311} (\bibinfo {year} {2019})}\BibitemShut {NoStop}%
\bibitem [{\citenamefont {Shi}\ \emph {et~al.}(2017)\citenamefont {Shi}, \citenamefont {Yu}, \citenamefont {Liu}, \citenamefont {He}, \citenamefont {Wang}, \citenamefont {Qin}, \citenamefont {Zhou}, \citenamefont {Li}, \citenamefont {Zhou}, \citenamefont {Sui}, \citenamefont {Zhang}, \citenamefont {Zhang}, \citenamefont {Zhang}, \citenamefont {Sum}, \citenamefont {Qiu}, \citenamefont {Liu},\ and\ \citenamefont {Liu}}]{shi3RMoS2Broken2017}%
  \BibitemOpen
  \bibfield  {author} {\bibinfo {author} {\bibfnamefont {J.}~\bibnamefont {Shi}}, \bibinfo {author} {\bibfnamefont {P.}~\bibnamefont {Yu}}, \bibinfo {author} {\bibfnamefont {F.}~\bibnamefont {Liu}}, \bibinfo {author} {\bibfnamefont {P.}~\bibnamefont {He}}, \bibinfo {author} {\bibfnamefont {R.}~\bibnamefont {Wang}}, \bibinfo {author} {\bibfnamefont {L.}~\bibnamefont {Qin}}, \bibinfo {author} {\bibfnamefont {J.}~\bibnamefont {Zhou}}, \bibinfo {author} {\bibfnamefont {X.}~\bibnamefont {Li}}, \bibinfo {author} {\bibfnamefont {J.}~\bibnamefont {Zhou}}, \bibinfo {author} {\bibfnamefont {X.}~\bibnamefont {Sui}}, \bibinfo {author} {\bibfnamefont {S.}~\bibnamefont {Zhang}}, \bibinfo {author} {\bibfnamefont {Y.}~\bibnamefont {Zhang}}, \bibinfo {author} {\bibfnamefont {Q.}~\bibnamefont {Zhang}}, \bibinfo {author} {\bibfnamefont {T.~C.}\ \bibnamefont {Sum}}, \bibinfo {author} {\bibfnamefont {X.}~\bibnamefont {Qiu}}, \bibinfo {author} {\bibfnamefont {Z.}~\bibnamefont {Liu}},\ and\ \bibinfo {author} {\bibfnamefont
  {X.}~\bibnamefont {Liu}},\ }\bibfield  {title} {\bibinfo {title} {{{3R MoS2}} with {{Broken Inversion Symmetry}}: {{A Promising Ultrathin Nonlinear Optical Device}}},\ }\href {https://doi.org/10.1002/adma.201701486} {\bibfield  {journal} {\bibinfo  {journal} {Advanced Materials}\ }\textbf {\bibinfo {volume} {29}},\ \bibinfo {pages} {1701486} (\bibinfo {year} {2017})}\BibitemShut {NoStop}%
\bibitem [{\citenamefont {Xu}\ \emph {et~al.}(2022)\citenamefont {Xu}, \citenamefont {Trovatello}, \citenamefont {Mooshammer}, \citenamefont {Shao}, \citenamefont {Zhang}, \citenamefont {Yao}, \citenamefont {Basov}, \citenamefont {Cerullo},\ and\ \citenamefont {Schuck}}]{xuCompactPhasematchedWaveguided2022}%
  \BibitemOpen
  \bibfield  {author} {\bibinfo {author} {\bibfnamefont {X.}~\bibnamefont {Xu}}, \bibinfo {author} {\bibfnamefont {C.}~\bibnamefont {Trovatello}}, \bibinfo {author} {\bibfnamefont {F.}~\bibnamefont {Mooshammer}}, \bibinfo {author} {\bibfnamefont {Y.}~\bibnamefont {Shao}}, \bibinfo {author} {\bibfnamefont {S.}~\bibnamefont {Zhang}}, \bibinfo {author} {\bibfnamefont {K.}~\bibnamefont {Yao}}, \bibinfo {author} {\bibfnamefont {D.~N.}\ \bibnamefont {Basov}}, \bibinfo {author} {\bibfnamefont {G.}~\bibnamefont {Cerullo}},\ and\ \bibinfo {author} {\bibfnamefont {P.~J.}\ \bibnamefont {Schuck}},\ }\bibfield  {title} {\bibinfo {title} {Towards compact phase-matched and waveguided nonlinear optics in atomically layered semiconductors},\ }\href {https://doi.org/10.1038/s41566-022-01053-4} {\bibfield  {journal} {\bibinfo  {journal} {Nature Photonics}\ }\textbf {\bibinfo {volume} {16}},\ \bibinfo {pages} {698} (\bibinfo {year} {2022})}\BibitemShut {NoStop}%
\bibitem [{\citenamefont {Wagoner}\ \emph {et~al.}(1998)\citenamefont {Wagoner}, \citenamefont {Persans}, \citenamefont {Van~Wagenen},\ and\ \citenamefont {Korenowski}}]{wagonerSecondharmonicGenerationMolybdenum1998}%
  \BibitemOpen
  \bibfield  {author} {\bibinfo {author} {\bibfnamefont {G.~A.}\ \bibnamefont {Wagoner}}, \bibinfo {author} {\bibfnamefont {P.~D.}\ \bibnamefont {Persans}}, \bibinfo {author} {\bibfnamefont {E.~A.}\ \bibnamefont {Van~Wagenen}},\ and\ \bibinfo {author} {\bibfnamefont {G.~M.}\ \bibnamefont {Korenowski}},\ }\bibfield  {title} {\bibinfo {title} {Second-harmonic generation in molybdenum disulfide},\ }\href {https://doi.org/10.1364/JOSAB.15.001017} {\bibfield  {journal} {\bibinfo  {journal} {Journal of the Optical Society of America B}\ }\textbf {\bibinfo {volume} {15}},\ \bibinfo {pages} {1017} (\bibinfo {year} {1998})}\BibitemShut {NoStop}%
\bibitem [{\citenamefont {Malard}\ \emph {et~al.}(2013)\citenamefont {Malard}, \citenamefont {Alencar}, \citenamefont {Barboza}, \citenamefont {Mak},\ and\ \citenamefont {de~Paula}}]{malardObservationIntenseSecond2013}%
  \BibitemOpen
  \bibfield  {author} {\bibinfo {author} {\bibfnamefont {L.~M.}\ \bibnamefont {Malard}}, \bibinfo {author} {\bibfnamefont {T.~V.}\ \bibnamefont {Alencar}}, \bibinfo {author} {\bibfnamefont {A.~P.~M.}\ \bibnamefont {Barboza}}, \bibinfo {author} {\bibfnamefont {K.~F.}\ \bibnamefont {Mak}},\ and\ \bibinfo {author} {\bibfnamefont {A.~M.}\ \bibnamefont {de~Paula}},\ }\bibfield  {title} {\bibinfo {title} {Observation of intense second harmonic generation from {{MoS}} 2 atomic crystals},\ }\href {https://doi.org/10.1103/PhysRevB.87.201401} {\bibfield  {journal} {\bibinfo  {journal} {Physical Review B}\ }\textbf {\bibinfo {volume} {87}},\ \bibinfo {pages} {201401(R)} (\bibinfo {year} {2013})}\BibitemShut {NoStop}%
\bibitem [{\citenamefont {Li}\ \emph {et~al.}(2013)\citenamefont {Li}, \citenamefont {Rao}, \citenamefont {Mak}, \citenamefont {You}, \citenamefont {Wang}, \citenamefont {Dean},\ and\ \citenamefont {Heinz}}]{liProbingSymmetryProperties2013}%
  \BibitemOpen
  \bibfield  {author} {\bibinfo {author} {\bibfnamefont {Y.}~\bibnamefont {Li}}, \bibinfo {author} {\bibfnamefont {Y.}~\bibnamefont {Rao}}, \bibinfo {author} {\bibfnamefont {K.~F.}\ \bibnamefont {Mak}}, \bibinfo {author} {\bibfnamefont {Y.}~\bibnamefont {You}}, \bibinfo {author} {\bibfnamefont {S.}~\bibnamefont {Wang}}, \bibinfo {author} {\bibfnamefont {C.~R.}\ \bibnamefont {Dean}},\ and\ \bibinfo {author} {\bibfnamefont {T.~F.}\ \bibnamefont {Heinz}},\ }\bibfield  {title} {\bibinfo {title} {Probing {{Symmetry Properties}} of {{Few-Layer MoS2}} and h-{{BN}} by {{Optical Second-Harmonic Generation}}},\ }\href {https://doi.org/10.1021/nl401561r} {\bibfield  {journal} {\bibinfo  {journal} {Nano Letters}\ }\textbf {\bibinfo {volume} {13}},\ \bibinfo {pages} {3329} (\bibinfo {year} {2013})}\BibitemShut {NoStop}%
\bibitem [{\citenamefont {Klimmer}\ \emph {et~al.}(2021)\citenamefont {Klimmer}, \citenamefont {Ghaebi}, \citenamefont {Gan}, \citenamefont {George}, \citenamefont {Turchanin}, \citenamefont {Cerullo},\ and\ \citenamefont {Soavi}}]{klimmerAllopticalPolarizationAmplitude2021}%
  \BibitemOpen
  \bibfield  {author} {\bibinfo {author} {\bibfnamefont {S.}~\bibnamefont {Klimmer}}, \bibinfo {author} {\bibfnamefont {O.}~\bibnamefont {Ghaebi}}, \bibinfo {author} {\bibfnamefont {Z.}~\bibnamefont {Gan}}, \bibinfo {author} {\bibfnamefont {A.}~\bibnamefont {George}}, \bibinfo {author} {\bibfnamefont {A.}~\bibnamefont {Turchanin}}, \bibinfo {author} {\bibfnamefont {G.}~\bibnamefont {Cerullo}},\ and\ \bibinfo {author} {\bibfnamefont {G.}~\bibnamefont {Soavi}},\ }\bibfield  {title} {\bibinfo {title} {All-optical polarization and amplitude modulation of second-harmonic generation in atomically thin semiconductors},\ }\href {https://doi.org/10.1038/s41566-021-00859-y} {\bibfield  {journal} {\bibinfo  {journal} {Nature Photonics}\ }\textbf {\bibinfo {volume} {15}},\ \bibinfo {pages} {837} (\bibinfo {year} {2021})}\BibitemShut {NoStop}%
\bibitem [{\citenamefont {Trovatello}\ \emph {et~al.}(2021)\citenamefont {Trovatello}, \citenamefont {Marini}, \citenamefont {Xu}, \citenamefont {Lee}, \citenamefont {Liu}, \citenamefont {Curreli}, \citenamefont {Manzoni}, \citenamefont {Dal~Conte}, \citenamefont {Yao}, \citenamefont {Ciattoni}, \citenamefont {Hone}, \citenamefont {Zhu}, \citenamefont {Schuck},\ and\ \citenamefont {Cerullo}}]{trovatelloOpticalParametricAmplification2021}%
  \BibitemOpen
  \bibfield  {author} {\bibinfo {author} {\bibfnamefont {C.}~\bibnamefont {Trovatello}}, \bibinfo {author} {\bibfnamefont {A.}~\bibnamefont {Marini}}, \bibinfo {author} {\bibfnamefont {X.}~\bibnamefont {Xu}}, \bibinfo {author} {\bibfnamefont {C.}~\bibnamefont {Lee}}, \bibinfo {author} {\bibfnamefont {F.}~\bibnamefont {Liu}}, \bibinfo {author} {\bibfnamefont {N.}~\bibnamefont {Curreli}}, \bibinfo {author} {\bibfnamefont {C.}~\bibnamefont {Manzoni}}, \bibinfo {author} {\bibfnamefont {S.}~\bibnamefont {Dal~Conte}}, \bibinfo {author} {\bibfnamefont {K.}~\bibnamefont {Yao}}, \bibinfo {author} {\bibfnamefont {A.}~\bibnamefont {Ciattoni}}, \bibinfo {author} {\bibfnamefont {J.}~\bibnamefont {Hone}}, \bibinfo {author} {\bibfnamefont {X.}~\bibnamefont {Zhu}}, \bibinfo {author} {\bibfnamefont {P.~J.}\ \bibnamefont {Schuck}},\ and\ \bibinfo {author} {\bibfnamefont {G.}~\bibnamefont {Cerullo}},\ }\bibfield  {title} {\bibinfo {title} {Optical parametric amplification by monolayer transition metal dichalcogenides},\ }\href
  {https://doi.org/10.1038/s41566-020-00728-0} {\bibfield  {journal} {\bibinfo  {journal} {Nature Photonics}\ }\textbf {\bibinfo {volume} {15}},\ \bibinfo {pages} {6} (\bibinfo {year} {2021})}\BibitemShut {NoStop}%
\bibitem [{\citenamefont {Hsu}\ \emph {et~al.}(2014)\citenamefont {Hsu}, \citenamefont {Zhao}, \citenamefont {Li}, \citenamefont {Chen}, \citenamefont {Chiu}, \citenamefont {Chang}, \citenamefont {Chou},\ and\ \citenamefont {Chang}}]{hsuSecondHarmonicGeneration2014}%
  \BibitemOpen
  \bibfield  {author} {\bibinfo {author} {\bibfnamefont {W.-T.}\ \bibnamefont {Hsu}}, \bibinfo {author} {\bibfnamefont {Z.-A.}\ \bibnamefont {Zhao}}, \bibinfo {author} {\bibfnamefont {L.-J.}\ \bibnamefont {Li}}, \bibinfo {author} {\bibfnamefont {C.-H.}\ \bibnamefont {Chen}}, \bibinfo {author} {\bibfnamefont {M.-H.}\ \bibnamefont {Chiu}}, \bibinfo {author} {\bibfnamefont {P.-S.}\ \bibnamefont {Chang}}, \bibinfo {author} {\bibfnamefont {Y.-C.}\ \bibnamefont {Chou}},\ and\ \bibinfo {author} {\bibfnamefont {W.-H.}\ \bibnamefont {Chang}},\ }\bibfield  {title} {\bibinfo {title} {Second {{Harmonic Generation}} from {{Artificially Stacked Transition Metal Dichalcogenide Twisted Bilayers}}},\ }\href {https://doi.org/10.1021/nn500228r} {\bibfield  {journal} {\bibinfo  {journal} {ACS Nano}\ }\textbf {\bibinfo {volume} {8}},\ \bibinfo {pages} {2951} (\bibinfo {year} {2014})}\BibitemShut {NoStop}%
\bibitem [{\citenamefont {Splendiani}\ \emph {et~al.}(2010)\citenamefont {Splendiani}, \citenamefont {Sun}, \citenamefont {Zhang}, \citenamefont {Li}, \citenamefont {Kim}, \citenamefont {Chim}, \citenamefont {Galli},\ and\ \citenamefont {Wang}}]{splendianiEmergingPhotoluminescenceMonolayer2010}%
  \BibitemOpen
  \bibfield  {author} {\bibinfo {author} {\bibfnamefont {A.}~\bibnamefont {Splendiani}}, \bibinfo {author} {\bibfnamefont {L.}~\bibnamefont {Sun}}, \bibinfo {author} {\bibfnamefont {Y.}~\bibnamefont {Zhang}}, \bibinfo {author} {\bibfnamefont {T.}~\bibnamefont {Li}}, \bibinfo {author} {\bibfnamefont {J.}~\bibnamefont {Kim}}, \bibinfo {author} {\bibfnamefont {C.-Y.}\ \bibnamefont {Chim}}, \bibinfo {author} {\bibfnamefont {G.}~\bibnamefont {Galli}},\ and\ \bibinfo {author} {\bibfnamefont {F.}~\bibnamefont {Wang}},\ }\bibfield  {title} {\bibinfo {title} {Emerging {{Photoluminescence}} in {{Monolayer MoS\textsubscript{2}}}},\ }\href {https://doi.org/10.1021/nl903868w} {\bibfield  {journal} {\bibinfo  {journal} {Nano Letters}\ }\textbf {\bibinfo {volume} {10}},\ \bibinfo {pages} {1271} (\bibinfo {year} {2010})}\BibitemShut {NoStop}%
\bibitem [{\citenamefont {Marini}\ \emph {et~al.}(2018)\citenamefont {Marini}, \citenamefont {Helt}, \citenamefont {Lu}, \citenamefont {Eggleton},\ and\ \citenamefont {Palomba}}]{mariniConstraintsDownconversionAtomically2018}%
  \BibitemOpen
  \bibfield  {author} {\bibinfo {author} {\bibfnamefont {L.}~\bibnamefont {Marini}}, \bibinfo {author} {\bibfnamefont {L.~G.}\ \bibnamefont {Helt}}, \bibinfo {author} {\bibfnamefont {Y.}~\bibnamefont {Lu}}, \bibinfo {author} {\bibfnamefont {B.~J.}\ \bibnamefont {Eggleton}},\ and\ \bibinfo {author} {\bibfnamefont {S.}~\bibnamefont {Palomba}},\ }\bibfield  {title} {\bibinfo {title} {Constraints on downconversion in atomically thick films},\ }\href {https://doi.org/10.1364/JOSAB.35.000672} {\bibfield  {journal} {\bibinfo  {journal} {Journal of the Optical Society of America B}\ }\textbf {\bibinfo {volume} {35}},\ \bibinfo {pages} {672} (\bibinfo {year} {2018})}\BibitemShut {NoStop}%
\bibitem [{\citenamefont {Dinparasti~Saleh}\ \emph {et~al.}(2018)\citenamefont {Dinparasti~Saleh}, \citenamefont {Vezzoli}, \citenamefont {Caspani}, \citenamefont {Branny}, \citenamefont {Kumar}, \citenamefont {Gerardot},\ and\ \citenamefont {Faccio}}]{dinparastisalehSpontaneousParametricConversion2018}%
  \BibitemOpen
  \bibfield  {author} {\bibinfo {author} {\bibfnamefont {H.}~\bibnamefont {Dinparasti~Saleh}}, \bibinfo {author} {\bibfnamefont {S.}~\bibnamefont {Vezzoli}}, \bibinfo {author} {\bibfnamefont {L.}~\bibnamefont {Caspani}}, \bibinfo {author} {\bibfnamefont {A.}~\bibnamefont {Branny}}, \bibinfo {author} {\bibfnamefont {S.}~\bibnamefont {Kumar}}, \bibinfo {author} {\bibfnamefont {B.~D.}\ \bibnamefont {Gerardot}},\ and\ \bibinfo {author} {\bibfnamefont {D.}~\bibnamefont {Faccio}},\ }\bibfield  {title} {\bibinfo {title} {Towards spontaneous parametric down conversion from monolayer {{MoS2}}},\ }\href {https://doi.org/10.1038/s41598-018-22270-4} {\bibfield  {journal} {\bibinfo  {journal} {Scientific Reports}\ }\textbf {\bibinfo {volume} {8}},\ \bibinfo {pages} {7842} (\bibinfo {year} {2018})}\BibitemShut {NoStop}%
\bibitem [{\citenamefont {Okoth}\ \emph {et~al.}(2019)\citenamefont {Okoth}, \citenamefont {Cavanna}, \citenamefont {Santiago-Cruz},\ and\ \citenamefont {Chekhova}}]{okothMicroscaleGenerationEntangled2019}%
  \BibitemOpen
  \bibfield  {author} {\bibinfo {author} {\bibfnamefont {C.}~\bibnamefont {Okoth}}, \bibinfo {author} {\bibfnamefont {A.}~\bibnamefont {Cavanna}}, \bibinfo {author} {\bibfnamefont {T.}~\bibnamefont {Santiago-Cruz}},\ and\ \bibinfo {author} {\bibfnamefont {M.~V.}\ \bibnamefont {Chekhova}},\ }\bibfield  {title} {\bibinfo {title} {Microscale {{Generation}} of {{Entangled Photons}} without {{Momentum Conservation}}},\ }\href {https://doi.org/10.1103/PhysRevLett.123.263602} {\bibfield  {journal} {\bibinfo  {journal} {Physical Review Letters}\ }\textbf {\bibinfo {volume} {123}},\ \bibinfo {pages} {263602} (\bibinfo {year} {2019})}\BibitemShut {NoStop}%
\bibitem [{\citenamefont {Sultanov}\ \emph {et~al.}(2022)\citenamefont {Sultanov}, \citenamefont {Santiago-Cruz},\ and\ \citenamefont {Chekhova}}]{sultanovFlatopticsGenerationBroadband2022}%
  \BibitemOpen
  \bibfield  {author} {\bibinfo {author} {\bibfnamefont {V.}~\bibnamefont {Sultanov}}, \bibinfo {author} {\bibfnamefont {T.}~\bibnamefont {Santiago-Cruz}},\ and\ \bibinfo {author} {\bibfnamefont {M.~V.}\ \bibnamefont {Chekhova}},\ }\bibfield  {title} {\bibinfo {title} {Flat-optics generation of broadband photon pairs with tunable polarization entanglement},\ }\href {https://doi.org/10.1364/OL.458133} {\bibfield  {journal} {\bibinfo  {journal} {Optics Letters}\ }\textbf {\bibinfo {volume} {47}},\ \bibinfo {pages} {3872} (\bibinfo {year} {2022})}\BibitemShut {NoStop}%
\bibitem [{\citenamefont {Guo}\ \emph {et~al.}(2023)\citenamefont {Guo}, \citenamefont {Qi}, \citenamefont {Zhang}, \citenamefont {Gao}, \citenamefont {Hu}, \citenamefont {Zhou}, \citenamefont {Zang}, \citenamefont {Zhao}, \citenamefont {Wang}, \citenamefont {Yan}, \citenamefont {Xu}, \citenamefont {Wu}, \citenamefont {Eda}, \citenamefont {Xiao}, \citenamefont {Yang}, \citenamefont {Gou}, \citenamefont {Feng}, \citenamefont {Guo}, \citenamefont {Zhou}, \citenamefont {Ren}, \citenamefont {Qiu}, \citenamefont {Pennycook},\ and\ \citenamefont {Wee}}]{guoUltrathinQuantumLight2023}%
  \BibitemOpen
  \bibfield  {author} {\bibinfo {author} {\bibfnamefont {Q.}~\bibnamefont {Guo}}, \bibinfo {author} {\bibfnamefont {X.-Z.}\ \bibnamefont {Qi}}, \bibinfo {author} {\bibfnamefont {L.}~\bibnamefont {Zhang}}, \bibinfo {author} {\bibfnamefont {M.}~\bibnamefont {Gao}}, \bibinfo {author} {\bibfnamefont {S.}~\bibnamefont {Hu}}, \bibinfo {author} {\bibfnamefont {W.}~\bibnamefont {Zhou}}, \bibinfo {author} {\bibfnamefont {W.}~\bibnamefont {Zang}}, \bibinfo {author} {\bibfnamefont {X.}~\bibnamefont {Zhao}}, \bibinfo {author} {\bibfnamefont {J.}~\bibnamefont {Wang}}, \bibinfo {author} {\bibfnamefont {B.}~\bibnamefont {Yan}}, \bibinfo {author} {\bibfnamefont {M.}~\bibnamefont {Xu}}, \bibinfo {author} {\bibfnamefont {Y.-K.}\ \bibnamefont {Wu}}, \bibinfo {author} {\bibfnamefont {G.}~\bibnamefont {Eda}}, \bibinfo {author} {\bibfnamefont {Z.}~\bibnamefont {Xiao}}, \bibinfo {author} {\bibfnamefont {S.~A.}\ \bibnamefont {Yang}}, \bibinfo {author} {\bibfnamefont {H.}~\bibnamefont {Gou}}, \bibinfo {author} {\bibfnamefont {Y.~P.}\
  \bibnamefont {Feng}}, \bibinfo {author} {\bibfnamefont {G.-C.}\ \bibnamefont {Guo}}, \bibinfo {author} {\bibfnamefont {W.}~\bibnamefont {Zhou}}, \bibinfo {author} {\bibfnamefont {X.-F.}\ \bibnamefont {Ren}}, \bibinfo {author} {\bibfnamefont {C.-W.}\ \bibnamefont {Qiu}}, \bibinfo {author} {\bibfnamefont {S.~J.}\ \bibnamefont {Pennycook}},\ and\ \bibinfo {author} {\bibfnamefont {A.~T.~S.}\ \bibnamefont {Wee}},\ }\bibfield  {title} {\bibinfo {title} {Ultrathin quantum light source with van der {Waals} {NbOCl2} crystal},\ }\href {https://doi.org/10.1038/s41586-022-05393-7} {\bibfield  {journal} {\bibinfo  {journal} {Nature}\ }\textbf {\bibinfo {volume} {613}},\ \bibinfo {pages} {53} (\bibinfo {year} {2023})}\BibitemShut {NoStop}%
\bibitem [{\citenamefont {Jeronimo-Moreno}\ \emph {et~al.}(2010)\citenamefont {Jeronimo-Moreno}, \citenamefont {Rodriguez-Benavides},\ and\ \citenamefont {U’Ren}}]{jeronimo-morenoTheoryCavityenhancedSpontaneous2010}%
  \BibitemOpen
  \bibfield  {author} {\bibinfo {author} {\bibfnamefont {Y.}~\bibnamefont {Jeronimo-Moreno}}, \bibinfo {author} {\bibfnamefont {S.}~\bibnamefont {Rodriguez-Benavides}},\ and\ \bibinfo {author} {\bibfnamefont {A.~B.}\ \bibnamefont {U’Ren}},\ }\bibfield  {title} {\bibinfo {title} {Theory of cavity-enhanced spontaneous parametric downconversion},\ }\href {https://doi.org/10.1134/S1054660X10090409} {\bibfield  {journal} {\bibinfo  {journal} {Laser Physics}\ }\textbf {\bibinfo {volume} {20}},\ \bibinfo {pages} {1221} (\bibinfo {year} {2010})}\BibitemShut {NoStop}%
\bibitem [{\citenamefont {Slattery}\ \emph {et~al.}(2019)\citenamefont {Slattery}, \citenamefont {Ma}, \citenamefont {Zong},\ and\ \citenamefont {Tang}}]{slatteryBackgroundReviewCavityEnhanced2019}%
  \BibitemOpen
  \bibfield  {author} {\bibinfo {author} {\bibfnamefont {O.}~\bibnamefont {Slattery}}, \bibinfo {author} {\bibfnamefont {L.}~\bibnamefont {Ma}}, \bibinfo {author} {\bibfnamefont {K.}~\bibnamefont {Zong}},\ and\ \bibinfo {author} {\bibfnamefont {X.}~\bibnamefont {Tang}},\ }\bibfield  {title} {\bibinfo {title} {Background and {{Review}} of {{Cavity-Enhanced Spontaneous Parametric Down-Conversion}}},\ }\href {https://doi.org/10.6028/jres.124.019} {\bibfield  {journal} {\bibinfo  {journal} {Journal of Research of the National Institute of Standards and Technology}\ }\textbf {\bibinfo {volume} {124}},\ \bibinfo {pages} {124019} (\bibinfo {year} {2019})}\BibitemShut {NoStop}%
\bibitem [{\citenamefont {Knopf}\ \emph {et~al.}(2019)\citenamefont {Knopf}, \citenamefont {Lundt}, \citenamefont {Bucher}, \citenamefont {Höfling}, \citenamefont {Tongay}, \citenamefont {Taniguchi}, \citenamefont {Watanabe}, \citenamefont {Staude}, \citenamefont {Schulz}, \citenamefont {Schneider},\ and\ \citenamefont {Eilenberger}}]{knopfIntegrationAtomicallyThin2019a}%
  \BibitemOpen
  \bibfield  {author} {\bibinfo {author} {\bibfnamefont {H.}~\bibnamefont {Knopf}}, \bibinfo {author} {\bibfnamefont {N.}~\bibnamefont {Lundt}}, \bibinfo {author} {\bibfnamefont {T.}~\bibnamefont {Bucher}}, \bibinfo {author} {\bibfnamefont {S.}~\bibnamefont {Höfling}}, \bibinfo {author} {\bibfnamefont {S.}~\bibnamefont {Tongay}}, \bibinfo {author} {\bibfnamefont {T.}~\bibnamefont {Taniguchi}}, \bibinfo {author} {\bibfnamefont {K.}~\bibnamefont {Watanabe}}, \bibinfo {author} {\bibfnamefont {I.}~\bibnamefont {Staude}}, \bibinfo {author} {\bibfnamefont {U.}~\bibnamefont {Schulz}}, \bibinfo {author} {\bibfnamefont {C.}~\bibnamefont {Schneider}},\ and\ \bibinfo {author} {\bibfnamefont {F.}~\bibnamefont {Eilenberger}},\ }\bibfield  {title} {\bibinfo {title} {Integration of atomically thin layers of transition metal dichalcogenides into high-{{Q}}, monolithic {{Bragg-cavities}}: An experimental platform for the enhancement of the optical interaction in {{2D-materials}}},\ }\href
  {https://doi.org/10.1364/OME.9.000598} {\bibfield  {journal} {\bibinfo  {journal} {Optical Materials Express}\ }\textbf {\bibinfo {volume} {9}},\ \bibinfo {pages} {598} (\bibinfo {year} {2019})}\BibitemShut {NoStop}%
\bibitem [{\citenamefont {Shradha}\ \emph {et~al.}()\citenamefont {Shradha}, \citenamefont {Abtahi}, \citenamefont {Gan}, \citenamefont {Knopf}, \citenamefont {Fedotova}, \citenamefont {Löchner}, \citenamefont {George}, \citenamefont {Pertsch}, \citenamefont {Turchanin},\ and\ \citenamefont {Eilenberger}}]{shradhaDoubleResonantCavity}%
  \BibitemOpen
  \bibfield  {author} {\bibinfo {author} {\bibfnamefont {S.}~\bibnamefont {Shradha}}, \bibinfo {author} {\bibfnamefont {F.}~\bibnamefont {Abtahi}}, \bibinfo {author} {\bibfnamefont {Z.}~\bibnamefont {Gan}}, \bibinfo {author} {\bibfnamefont {H.}~\bibnamefont {Knopf}}, \bibinfo {author} {\bibfnamefont {A.}~\bibnamefont {Fedotova}}, \bibinfo {author} {\bibfnamefont {F.~J.~F.}\ \bibnamefont {Löchner}}, \bibinfo {author} {\bibfnamefont {A.}~\bibnamefont {George}}, \bibinfo {author} {\bibfnamefont {T.}~\bibnamefont {Pertsch}}, \bibinfo {author} {\bibfnamefont {A.}~\bibnamefont {Turchanin}},\ and\ \bibinfo {author} {\bibfnamefont {F.}~\bibnamefont {Eilenberger}},\ }\bibfield  {title} {\bibinfo {title} {Towards {{Double Resonant Cavity Enhanced Second Harmonic Generation}} in {{Monolayer MoS2}}},\ }\href {https://doi.org/10.1002/adom.202300907} {\bibfield  {journal} {\bibinfo  {journal} {Advanced Optical Materials}\ }\textbf {\bibinfo {volume} {n/a}},\ \bibinfo {pages} {2300907}}\BibitemShut {NoStop}%
\bibitem [{\citenamefont {Boyd}(2008)}]{boydNonlinearOptics2008}%
  \BibitemOpen
  \bibfield  {author} {\bibinfo {author} {\bibfnamefont {R.~W.}\ \bibnamefont {Boyd}},\ }\href@noop {} {\emph {\bibinfo {title} {Nonlinear {{Optics}}}}},\ \bibinfo {edition} {3rd}\ ed.\ (\bibinfo  {publisher} {{Academic Press}},\ \bibinfo {year} {2008})\BibitemShut {NoStop}%
\bibitem [{\citenamefont {Wootters}(1998)}]{woottersEntanglementFormationArbitrary1998}%
  \BibitemOpen
  \bibfield  {author} {\bibinfo {author} {\bibfnamefont {W.~K.}\ \bibnamefont {Wootters}},\ }\bibfield  {title} {\bibinfo {title} {Entanglement of {{Formation}} of an {{Arbitrary State}} of {{Two Qubits}}},\ }\href {https://doi.org/10.1103/PhysRevLett.80.2245} {\bibfield  {journal} {\bibinfo  {journal} {Physical Review Letters}\ }\textbf {\bibinfo {volume} {80}},\ \bibinfo {pages} {2245} (\bibinfo {year} {1998})}\BibitemShut {NoStop}%
\bibitem [{\citenamefont {Mennel}\ \emph {et~al.}(2019)\citenamefont {Mennel}, \citenamefont {Paur},\ and\ \citenamefont {Mueller}}]{mennelSecondHarmonicGeneration2019}%
  \BibitemOpen
  \bibfield  {author} {\bibinfo {author} {\bibfnamefont {L.}~\bibnamefont {Mennel}}, \bibinfo {author} {\bibfnamefont {M.}~\bibnamefont {Paur}},\ and\ \bibinfo {author} {\bibfnamefont {T.}~\bibnamefont {Mueller}},\ }\bibfield  {title} {\bibinfo {title} {Second harmonic generation in strained transition metal dichalcogenide monolayers: {{MoS2}}, {{MoSe2}}, {{WS2}}, and {{WSe2}}},\ }\href {https://doi.org/10.1063/1.5051965} {\bibfield  {journal} {\bibinfo  {journal} {APL Photonics}\ }\textbf {\bibinfo {volume} {4}},\ \bibinfo {pages} {034404} (\bibinfo {year} {2019})}\BibitemShut {NoStop}%
\bibitem [{\citenamefont {Valencia}\ \emph {et~al.}(2002)\citenamefont {Valencia}, \citenamefont {Chekhova}, \citenamefont {Trifonov},\ and\ \citenamefont {Shih}}]{valenciaEntangledTwoPhotonWave2002a}%
  \BibitemOpen
  \bibfield  {author} {\bibinfo {author} {\bibfnamefont {A.}~\bibnamefont {Valencia}}, \bibinfo {author} {\bibfnamefont {M.~V.}\ \bibnamefont {Chekhova}}, \bibinfo {author} {\bibfnamefont {A.}~\bibnamefont {Trifonov}},\ and\ \bibinfo {author} {\bibfnamefont {Y.}~\bibnamefont {Shih}},\ }\bibfield  {title} {\bibinfo {title} {Entangled {{Two-Photon Wave Packet}} in a {{Dispersive Medium}}},\ }\href {https://doi.org/10.1103/PhysRevLett.88.183601} {\bibfield  {journal} {\bibinfo  {journal} {Physical Review Letters}\ }\textbf {\bibinfo {volume} {88}},\ \bibinfo {pages} {183601} (\bibinfo {year} {2002})}\BibitemShut {NoStop}%
\bibitem [{\citenamefont {Santiago-Cruz}\ \emph {et~al.}(2021{\natexlab{a}})\citenamefont {Santiago-Cruz}, \citenamefont {Santiago-Cruz}, \citenamefont {Sultanov}, \citenamefont {Sultanov}, \citenamefont {Zhang}, \citenamefont {Krivitsky}, \citenamefont {Chekhova},\ and\ \citenamefont {Chekhova}}]{santiago-cruzEntangledPhotonsSubwavelength2021}%
  \BibitemOpen
  \bibfield  {author} {\bibinfo {author} {\bibfnamefont {T.}~\bibnamefont {Santiago-Cruz}}, \bibinfo {author} {\bibfnamefont {T.}~\bibnamefont {Santiago-Cruz}}, \bibinfo {author} {\bibfnamefont {V.}~\bibnamefont {Sultanov}}, \bibinfo {author} {\bibfnamefont {V.}~\bibnamefont {Sultanov}}, \bibinfo {author} {\bibfnamefont {H.}~\bibnamefont {Zhang}}, \bibinfo {author} {\bibfnamefont {L.~A.}\ \bibnamefont {Krivitsky}}, \bibinfo {author} {\bibfnamefont {M.~V.}\ \bibnamefont {Chekhova}},\ and\ \bibinfo {author} {\bibfnamefont {M.~V.}\ \bibnamefont {Chekhova}},\ }\bibfield  {title} {\bibinfo {title} {Entangled photons from subwavelength nonlinear films},\ }\href {https://doi.org/10.1364/OL.411176} {\bibfield  {journal} {\bibinfo  {journal} {Optics Letters}\ }\textbf {\bibinfo {volume} {46}},\ \bibinfo {pages} {653} (\bibinfo {year} {2021}{\natexlab{a}})}\BibitemShut {NoStop}%
\bibitem [{\citenamefont {Chang}\ \emph {et~al.}(2019)\citenamefont {Chang}, \citenamefont {Zadeh}, \citenamefont {Los}, \citenamefont {Zichi}, \citenamefont {Fognini}, \citenamefont {Gevers}, \citenamefont {Dorenbos}, \citenamefont {Pereira}, \citenamefont {Urbach},\ and\ \citenamefont {Zwiller}}]{changMultimodefibercoupledSuperconductingNanowire2019}%
  \BibitemOpen
  \bibfield  {author} {\bibinfo {author} {\bibfnamefont {J.}~\bibnamefont {Chang}}, \bibinfo {author} {\bibfnamefont {I.~E.}\ \bibnamefont {Zadeh}}, \bibinfo {author} {\bibfnamefont {J.~W.~N.}\ \bibnamefont {Los}}, \bibinfo {author} {\bibfnamefont {J.}~\bibnamefont {Zichi}}, \bibinfo {author} {\bibfnamefont {A.}~\bibnamefont {Fognini}}, \bibinfo {author} {\bibfnamefont {M.}~\bibnamefont {Gevers}}, \bibinfo {author} {\bibfnamefont {S.}~\bibnamefont {Dorenbos}}, \bibinfo {author} {\bibfnamefont {S.~F.}\ \bibnamefont {Pereira}}, \bibinfo {author} {\bibfnamefont {P.}~\bibnamefont {Urbach}},\ and\ \bibinfo {author} {\bibfnamefont {V.}~\bibnamefont {Zwiller}},\ }\bibfield  {title} {\bibinfo {title} {Multimode-fiber-coupled superconducting nanowire single-photon detectors with high detection efficiency and time resolution},\ }\href {https://doi.org/10.1364/AO.58.009803} {\bibfield  {journal} {\bibinfo  {journal} {Applied Optics}\ }\textbf {\bibinfo {volume} {58}},\ \bibinfo {pages} {9803} (\bibinfo {year}
  {2019})}\BibitemShut {NoStop}%
\bibitem [{\citenamefont {James}\ \emph {et~al.}(2001)\citenamefont {James}, \citenamefont {Kwiat}, \citenamefont {Munro},\ and\ \citenamefont {White}}]{jamesMeasurementQubits2001}%
  \BibitemOpen
  \bibfield  {author} {\bibinfo {author} {\bibfnamefont {D.~F.~V.}\ \bibnamefont {James}}, \bibinfo {author} {\bibfnamefont {P.~G.}\ \bibnamefont {Kwiat}}, \bibinfo {author} {\bibfnamefont {W.~J.}\ \bibnamefont {Munro}},\ and\ \bibinfo {author} {\bibfnamefont {A.~G.}\ \bibnamefont {White}},\ }\bibfield  {title} {\bibinfo {title} {Measurement of qubits},\ }\href {https://doi.org/10.1103/PhysRevA.64.052312} {\bibfield  {journal} {\bibinfo  {journal} {Physical Review A}\ }\textbf {\bibinfo {volume} {64}},\ \bibinfo {pages} {052312} (\bibinfo {year} {2001})}\BibitemShut {NoStop}%
\bibitem [{Qua()}]{QuantumTomography2023}%
  \BibitemOpen
  \href {https://github.com/KwiatLab/Quantum-Tomography} {\bibinfo {title} {Quantum-{{Tomography}}}},\ \bibinfo {howpublished} {https://github.com/KwiatLab/\\Quantum-Tomography},\ \bibinfo {note} {[Accessed: 2023-08-16]}\BibitemShut {NoStop}%
\bibitem [{\citenamefont {Xuan}\ \emph {et~al.}(2023)\citenamefont {Xuan}, \citenamefont {Lai}, \citenamefont {Wu},\ and\ \citenamefont {Quek}}]{xuanExcitonenhancedSpontaneousParametric2023}%
  \BibitemOpen
  \bibfield  {author} {\bibinfo {author} {\bibfnamefont {F.}~\bibnamefont {Xuan}}, \bibinfo {author} {\bibfnamefont {M.}~\bibnamefont {Lai}}, \bibinfo {author} {\bibfnamefont {Y.}~\bibnamefont {Wu}},\ and\ \bibinfo {author} {\bibfnamefont {S.~Y.}\ \bibnamefont {Quek}},\ }\bibfield  {title} {\bibinfo {title} {Exciton-enhanced {{Spontaneous Parametric Down-Conversion}} in {{2D Crystals}}},\ }\Eprint {https://arxiv.org/abs/2305.08345} {arXiv:2305.08345 [physics:comp-ph, cond-mat:mtrl-sci]}  (\bibinfo {year} {2023})\BibitemShut {NoStop}%
\bibitem [{\citenamefont {Nauman}\ \emph {et~al.}(2021)\citenamefont {Nauman}, \citenamefont {Yan}, \citenamefont {Ceglia}, \citenamefont {Rahmani}, \citenamefont {Zangeneh~Kamali}, \citenamefont {De~Angelis}, \citenamefont {Miroshnichenko}, \citenamefont {Lu},\ and\ \citenamefont {Neshev}}]{naumanTunableUnidirectionalNonlinear2021}%
  \BibitemOpen
  \bibfield  {author} {\bibinfo {author} {\bibfnamefont {M.}~\bibnamefont {Nauman}}, \bibinfo {author} {\bibfnamefont {J.}~\bibnamefont {Yan}}, \bibinfo {author} {\bibfnamefont {D.~d.}\ \bibnamefont {Ceglia}}, \bibinfo {author} {\bibfnamefont {M.}~\bibnamefont {Rahmani}}, \bibinfo {author} {\bibfnamefont {K.}~\bibnamefont {Zangeneh~Kamali}}, \bibinfo {author} {\bibfnamefont {C.}~\bibnamefont {De~Angelis}}, \bibinfo {author} {\bibfnamefont {A.~E.}\ \bibnamefont {Miroshnichenko}}, \bibinfo {author} {\bibfnamefont {Y.}~\bibnamefont {Lu}},\ and\ \bibinfo {author} {\bibfnamefont {D.~N.}\ \bibnamefont {Neshev}},\ }\bibfield  {title} {\bibinfo {title} {Tunable unidirectional nonlinear emission from transition-metal-dichalcogenide metasurfaces},\ }\href {https://doi.org/10.1038/s41467-021-25717-x} {\bibfield  {journal} {\bibinfo  {journal} {Nature Communications}\ }\textbf {\bibinfo {volume} {12}},\ \bibinfo {pages} {5597} (\bibinfo {year} {2021})}\BibitemShut {NoStop}%
\bibitem [{\citenamefont {Zotev}\ \emph {et~al.}(2022)\citenamefont {Zotev}, \citenamefont {Wang}, \citenamefont {Sortino}, \citenamefont {Severs~Millard}, \citenamefont {Mullin}, \citenamefont {Conteduca}, \citenamefont {Shagar}, \citenamefont {Genco}, \citenamefont {Hobbs}, \citenamefont {Krauss},\ and\ \citenamefont {Tartakovskii}}]{zotevTransitionMetalDichalcogenide2022}%
  \BibitemOpen
  \bibfield  {author} {\bibinfo {author} {\bibfnamefont {P.~G.}\ \bibnamefont {Zotev}}, \bibinfo {author} {\bibfnamefont {Y.}~\bibnamefont {Wang}}, \bibinfo {author} {\bibfnamefont {L.}~\bibnamefont {Sortino}}, \bibinfo {author} {\bibfnamefont {T.}~\bibnamefont {Severs~Millard}}, \bibinfo {author} {\bibfnamefont {N.}~\bibnamefont {Mullin}}, \bibinfo {author} {\bibfnamefont {D.}~\bibnamefont {Conteduca}}, \bibinfo {author} {\bibfnamefont {M.}~\bibnamefont {Shagar}}, \bibinfo {author} {\bibfnamefont {A.}~\bibnamefont {Genco}}, \bibinfo {author} {\bibfnamefont {J.~K.}\ \bibnamefont {Hobbs}}, \bibinfo {author} {\bibfnamefont {T.~F.}\ \bibnamefont {Krauss}},\ and\ \bibinfo {author} {\bibfnamefont {A.~I.}\ \bibnamefont {Tartakovskii}},\ }\bibfield  {title} {\bibinfo {title} {Transition {{Metal Dichalcogenide Dimer Nanoantennas}} for {{Tailored Light}}–{{Matter Interactions}}},\ }\href {https://doi.org/10.1021/acsnano.2c00802} {\bibfield  {journal} {\bibinfo  {journal} {ACS Nano}\ }\textbf {\bibinfo {volume}
  {16}},\ \bibinfo {pages} {6493} (\bibinfo {year} {2022})}\BibitemShut {NoStop}%
\bibitem [{\citenamefont {Santiago-Cruz}\ \emph {et~al.}(2021{\natexlab{b}})\citenamefont {Santiago-Cruz}, \citenamefont {Fedotova}, \citenamefont {Sultanov}, \citenamefont {Weissflog}, \citenamefont {Arslan}, \citenamefont {Younesi}, \citenamefont {Pertsch}, \citenamefont {Staude}, \citenamefont {Setzpfandt},\ and\ \citenamefont {Chekhova}}]{santiago-cruzPhotonPairsResonant2021}%
  \BibitemOpen
  \bibfield  {author} {\bibinfo {author} {\bibfnamefont {T.}~\bibnamefont {Santiago-Cruz}}, \bibinfo {author} {\bibfnamefont {A.}~\bibnamefont {Fedotova}}, \bibinfo {author} {\bibfnamefont {V.}~\bibnamefont {Sultanov}}, \bibinfo {author} {\bibfnamefont {M.~A.}\ \bibnamefont {Weissflog}}, \bibinfo {author} {\bibfnamefont {D.}~\bibnamefont {Arslan}}, \bibinfo {author} {\bibfnamefont {M.}~\bibnamefont {Younesi}}, \bibinfo {author} {\bibfnamefont {T.}~\bibnamefont {Pertsch}}, \bibinfo {author} {\bibfnamefont {I.}~\bibnamefont {Staude}}, \bibinfo {author} {\bibfnamefont {F.}~\bibnamefont {Setzpfandt}},\ and\ \bibinfo {author} {\bibfnamefont {M.}~\bibnamefont {Chekhova}},\ }\bibfield  {title} {\bibinfo {title} {Photon {{Pairs}} from {{Resonant Metasurfaces}}},\ }\href {https://doi.org/10.1021/acs.nanolett.1c01125} {\bibfield  {journal} {\bibinfo  {journal} {Nano Letters}\ }\textbf {\bibinfo {volume} {21}},\ \bibinfo {pages} {4423} (\bibinfo {year} {2021}{\natexlab{b}})}\BibitemShut {NoStop}%
\bibitem [{\citenamefont {Santiago-Cruz}\ \emph {et~al.}(2022)\citenamefont {Santiago-Cruz}, \citenamefont {Gennaro}, \citenamefont {Mitrofanov}, \citenamefont {Addamane}, \citenamefont {Reno}, \citenamefont {Brener},\ and\ \citenamefont {Chekhova}}]{santiago-cruzResonantMetasurfacesGenerating2022}%
  \BibitemOpen
  \bibfield  {author} {\bibinfo {author} {\bibfnamefont {T.}~\bibnamefont {Santiago-Cruz}}, \bibinfo {author} {\bibfnamefont {S.~D.}\ \bibnamefont {Gennaro}}, \bibinfo {author} {\bibfnamefont {O.}~\bibnamefont {Mitrofanov}}, \bibinfo {author} {\bibfnamefont {S.}~\bibnamefont {Addamane}}, \bibinfo {author} {\bibfnamefont {J.}~\bibnamefont {Reno}}, \bibinfo {author} {\bibfnamefont {I.}~\bibnamefont {Brener}},\ and\ \bibinfo {author} {\bibfnamefont {M.~V.}\ \bibnamefont {Chekhova}},\ }\bibfield  {title} {\bibinfo {title} {Resonant metasurfaces for generating complex quantum states},\ }\href {https://doi.org/10.1126/science.abq8684} {\bibfield  {journal} {\bibinfo  {journal} {Science}\ }\textbf {\bibinfo {volume} {377}},\ \bibinfo {pages} {991} (\bibinfo {year} {2022})}\BibitemShut {NoStop}%
\bibitem [{\citenamefont {Weissflog}\ \emph {et~al.}(2023)\citenamefont {Weissflog}, \citenamefont {Dezert}, \citenamefont {Vinel}, \citenamefont {Gigli}, \citenamefont {Leo}, \citenamefont {Pertsch}, \citenamefont {Setzpfandt}, \citenamefont {Borne},\ and\ \citenamefont {Saravi}}]{weissflog_nonlinear_2023}%
  \BibitemOpen
  \bibfield  {author} {\bibinfo {author} {\bibfnamefont {M.~A.}\ \bibnamefont {Weissflog}}, \bibinfo {author} {\bibfnamefont {R.}~\bibnamefont {Dezert}}, \bibinfo {author} {\bibfnamefont {V.}~\bibnamefont {Vinel}}, \bibinfo {author} {\bibfnamefont {C.}~\bibnamefont {Gigli}}, \bibinfo {author} {\bibfnamefont {G.}~\bibnamefont {Leo}}, \bibinfo {author} {\bibfnamefont {T.}~\bibnamefont {Pertsch}}, \bibinfo {author} {\bibfnamefont {F.}~\bibnamefont {Setzpfandt}}, \bibinfo {author} {\bibfnamefont {A.}~\bibnamefont {Borne}},\ and\ \bibinfo {author} {\bibfnamefont {S.}~\bibnamefont {Saravi}},\ }\bibfield  {title} {\bibinfo {title} {Nonlinear nanoresonators are natural sources of bell states},\ }\Eprint {https://arxiv.org/abs/2305.19362} {arXiv:2305.19362 [physics:optics]}  (\bibinfo {year} {2023})\BibitemShut {NoStop}%
\bibitem [{\citenamefont {Vogl}\ \emph {et~al.}(2019)\citenamefont {Vogl}, \citenamefont {Sripathy}, \citenamefont {Sharma}, \citenamefont {Reddy}, \citenamefont {Sullivan}, \citenamefont {Machacek}, \citenamefont {Zhang}, \citenamefont {Karouta}, \citenamefont {Buchler}, \citenamefont {Doherty}, \citenamefont {Lu},\ and\ \citenamefont {Lam}}]{voglRadiationToleranceTwodimensional2019}%
  \BibitemOpen
  \bibfield  {author} {\bibinfo {author} {\bibfnamefont {T.}~\bibnamefont {Vogl}}, \bibinfo {author} {\bibfnamefont {K.}~\bibnamefont {Sripathy}}, \bibinfo {author} {\bibfnamefont {A.}~\bibnamefont {Sharma}}, \bibinfo {author} {\bibfnamefont {P.}~\bibnamefont {Reddy}}, \bibinfo {author} {\bibfnamefont {J.}~\bibnamefont {Sullivan}}, \bibinfo {author} {\bibfnamefont {J.~R.}\ \bibnamefont {Machacek}}, \bibinfo {author} {\bibfnamefont {L.}~\bibnamefont {Zhang}}, \bibinfo {author} {\bibfnamefont {F.}~\bibnamefont {Karouta}}, \bibinfo {author} {\bibfnamefont {B.~C.}\ \bibnamefont {Buchler}}, \bibinfo {author} {\bibfnamefont {M.~W.}\ \bibnamefont {Doherty}}, \bibinfo {author} {\bibfnamefont {Y.}~\bibnamefont {Lu}},\ and\ \bibinfo {author} {\bibfnamefont {P.~K.}\ \bibnamefont {Lam}},\ }\bibfield  {title} {\bibinfo {title} {Radiation tolerance of two-dimensional material-based devices for space applications},\ }\href {https://doi.org/10.1038/s41467-019-09219-5} {\bibfield  {journal} {\bibinfo  {journal} {Nature
  Communications}\ }\textbf {\bibinfo {volume} {10}},\ \bibinfo {pages} {1202} (\bibinfo {year} {2019})}\BibitemShut {NoStop}%
\bibitem [{\citenamefont {Ngo}\ \emph {et~al.}(2022)\citenamefont {Ngo}, \citenamefont {Najafidehaghani}, \citenamefont {Gan}, \citenamefont {Khazaee}, \citenamefont {Siems}, \citenamefont {George}, \citenamefont {Schartner}, \citenamefont {Nolte}, \citenamefont {Ebendorff-Heidepriem}, \citenamefont {Pertsch}, \citenamefont {Tuniz}, \citenamefont {Schmidt}, \citenamefont {Peschel}, \citenamefont {Turchanin},\ and\ \citenamefont {Eilenberger}}]{ngoInfibreSecondharmonicGeneration2022}%
  \BibitemOpen
  \bibfield  {author} {\bibinfo {author} {\bibfnamefont {G.~Q.}\ \bibnamefont {Ngo}}, \bibinfo {author} {\bibfnamefont {E.}~\bibnamefont {Najafidehaghani}}, \bibinfo {author} {\bibfnamefont {Z.}~\bibnamefont {Gan}}, \bibinfo {author} {\bibfnamefont {S.}~\bibnamefont {Khazaee}}, \bibinfo {author} {\bibfnamefont {M.~P.}\ \bibnamefont {Siems}}, \bibinfo {author} {\bibfnamefont {A.}~\bibnamefont {George}}, \bibinfo {author} {\bibfnamefont {E.~P.}\ \bibnamefont {Schartner}}, \bibinfo {author} {\bibfnamefont {S.}~\bibnamefont {Nolte}}, \bibinfo {author} {\bibfnamefont {H.}~\bibnamefont {Ebendorff-Heidepriem}}, \bibinfo {author} {\bibfnamefont {T.}~\bibnamefont {Pertsch}}, \bibinfo {author} {\bibfnamefont {A.}~\bibnamefont {Tuniz}}, \bibinfo {author} {\bibfnamefont {M.~A.}\ \bibnamefont {Schmidt}}, \bibinfo {author} {\bibfnamefont {U.}~\bibnamefont {Peschel}}, \bibinfo {author} {\bibfnamefont {A.}~\bibnamefont {Turchanin}},\ and\ \bibinfo {author} {\bibfnamefont {F.}~\bibnamefont {Eilenberger}},\ }\bibfield  {title}
  {\bibinfo {title} {In-fibre second-harmonic generation with embedded two-dimensional materials},\ }\href {https://doi.org/10.1038/s41566-022-01067-y} {\bibfield  {journal} {\bibinfo  {journal} {Nature Photonics}\ }\textbf {\bibinfo {volume} {16}},\ \bibinfo {pages} {769} (\bibinfo {year} {2022})}\BibitemShut {NoStop}%
\bibitem [{\citenamefont {Vogl}\ \emph {et~al.}(2017)\citenamefont {Vogl}, \citenamefont {Lu},\ and\ \citenamefont {Lam}}]{voglRoomTemperatureSingle2017a}%
  \BibitemOpen
  \bibfield  {author} {\bibinfo {author} {\bibfnamefont {T.}~\bibnamefont {Vogl}}, \bibinfo {author} {\bibfnamefont {Y.}~\bibnamefont {Lu}},\ and\ \bibinfo {author} {\bibfnamefont {P.~K.}\ \bibnamefont {Lam}},\ }\bibfield  {title} {\bibinfo {title} {Room temperature single photon source using fiber-integrated hexagonal boron nitride},\ }\href {https://doi.org/10.1088/1361-6463/aa7839} {\bibfield  {journal} {\bibinfo  {journal} {Journal of Physics D: Applied Physics}\ }\textbf {\bibinfo {volume} {50}},\ \bibinfo {pages} {295101} (\bibinfo {year} {2017})}\BibitemShut {NoStop}%
\bibitem [{\citenamefont {Kuppadakkath}\ \emph {et~al.}(2022)\citenamefont {Kuppadakkath}, \citenamefont {Najafidehaghani}, \citenamefont {Gan}, \citenamefont {Tuniz}, \citenamefont {Ngo}, \citenamefont {Knopf}, \citenamefont {Löchner}, \citenamefont {Abtahi}, \citenamefont {Bucher}, \citenamefont {Shradha}, \citenamefont {Käsebier}, \citenamefont {Palomba}, \citenamefont {Felde}, \citenamefont {Paul}, \citenamefont {Ullsperger}, \citenamefont {Schröder}, \citenamefont {Szeghalmi}, \citenamefont {Pertsch}, \citenamefont {Staude}, \citenamefont {Zeitner}, \citenamefont {George}, \citenamefont {Turchanin},\ and\ \citenamefont {Eilenberger}}]{kuppadakkathDirectGrowthMonolayer2022}%
  \BibitemOpen
  \bibfield  {author} {\bibinfo {author} {\bibfnamefont {A.}~\bibnamefont {Kuppadakkath}}, \bibinfo {author} {\bibfnamefont {E.}~\bibnamefont {Najafidehaghani}}, \bibinfo {author} {\bibfnamefont {Z.}~\bibnamefont {Gan}}, \bibinfo {author} {\bibfnamefont {A.}~\bibnamefont {Tuniz}}, \bibinfo {author} {\bibfnamefont {G.~Q.}\ \bibnamefont {Ngo}}, \bibinfo {author} {\bibfnamefont {H.}~\bibnamefont {Knopf}}, \bibinfo {author} {\bibfnamefont {F.~J.~F.}\ \bibnamefont {Löchner}}, \bibinfo {author} {\bibfnamefont {F.}~\bibnamefont {Abtahi}}, \bibinfo {author} {\bibfnamefont {T.}~\bibnamefont {Bucher}}, \bibinfo {author} {\bibfnamefont {S.}~\bibnamefont {Shradha}}, \bibinfo {author} {\bibfnamefont {T.}~\bibnamefont {Käsebier}}, \bibinfo {author} {\bibfnamefont {S.}~\bibnamefont {Palomba}}, \bibinfo {author} {\bibfnamefont {N.}~\bibnamefont {Felde}}, \bibinfo {author} {\bibfnamefont {P.}~\bibnamefont {Paul}}, \bibinfo {author} {\bibfnamefont {T.}~\bibnamefont {Ullsperger}}, \bibinfo {author} {\bibfnamefont
  {S.}~\bibnamefont {Schröder}}, \bibinfo {author} {\bibfnamefont {A.}~\bibnamefont {Szeghalmi}}, \bibinfo {author} {\bibfnamefont {T.}~\bibnamefont {Pertsch}}, \bibinfo {author} {\bibfnamefont {I.}~\bibnamefont {Staude}}, \bibinfo {author} {\bibfnamefont {U.}~\bibnamefont {Zeitner}}, \bibinfo {author} {\bibfnamefont {A.}~\bibnamefont {George}}, \bibinfo {author} {\bibfnamefont {A.}~\bibnamefont {Turchanin}},\ and\ \bibinfo {author} {\bibfnamefont {F.}~\bibnamefont {Eilenberger}},\ }\bibfield  {title} {\bibinfo {title} {Direct growth of monolayer {{MoS2}} on nanostructured silicon waveguides},\ }\href {https://doi.org/10.1515/nanoph-2022-0235} {\bibfield  {journal} {\bibinfo  {journal} {Nanophotonics}\ }\textbf {\bibinfo {volume} {11}},\ \bibinfo {pages} {4397} (\bibinfo {year} {2022})}\BibitemShut {NoStop}%
\bibitem [{\citenamefont {Löchner}\ \emph {et~al.}(2021)\citenamefont {Löchner}, \citenamefont {George}, \citenamefont {Koshelev}, \citenamefont {Bucher}, \citenamefont {Najafidehaghani}, \citenamefont {Fedotova}, \citenamefont {Choi}, \citenamefont {Pertsch}, \citenamefont {Staude}, \citenamefont {Kivshar}, \citenamefont {Turchanin},\ and\ \citenamefont {Setzpfandt}}]{lochnerHybridDielectricMetasurfaces2021}%
  \BibitemOpen
  \bibfield  {author} {\bibinfo {author} {\bibfnamefont {F.~J.~F.}\ \bibnamefont {Löchner}}, \bibinfo {author} {\bibfnamefont {A.}~\bibnamefont {George}}, \bibinfo {author} {\bibfnamefont {K.}~\bibnamefont {Koshelev}}, \bibinfo {author} {\bibfnamefont {T.}~\bibnamefont {Bucher}}, \bibinfo {author} {\bibfnamefont {E.}~\bibnamefont {Najafidehaghani}}, \bibinfo {author} {\bibfnamefont {A.}~\bibnamefont {Fedotova}}, \bibinfo {author} {\bibfnamefont {D.-Y.}\ \bibnamefont {Choi}}, \bibinfo {author} {\bibfnamefont {T.}~\bibnamefont {Pertsch}}, \bibinfo {author} {\bibfnamefont {I.}~\bibnamefont {Staude}}, \bibinfo {author} {\bibfnamefont {Y.}~\bibnamefont {Kivshar}}, \bibinfo {author} {\bibfnamefont {A.}~\bibnamefont {Turchanin}},\ and\ \bibinfo {author} {\bibfnamefont {F.}~\bibnamefont {Setzpfandt}},\ }\bibfield  {title} {\bibinfo {title} {Hybrid {{Dielectric Metasurfaces}} for {{Enhancing Second-Harmonic Generation}} in {{Chemical Vapor Deposition Grown MoS2 Monolayers}}},\ }\href
  {https://doi.org/10.1021/acsphotonics.0c01375} {\bibfield  {journal} {\bibinfo  {journal} {ACS Photonics}\ }\textbf {\bibinfo {volume} {8}},\ \bibinfo {pages} {218} (\bibinfo {year} {2021})}\BibitemShut {NoStop}%
\bibitem [{\citenamefont {Altepeter}\ \emph {et~al.}(2004)\citenamefont {Altepeter}, \citenamefont {James},\ and\ \citenamefont {Kwiat}}]{altepeterQubitQuantumState2004}%
  \BibitemOpen
  \bibfield  {author} {\bibinfo {author} {\bibfnamefont {J.~B.}\ \bibnamefont {Altepeter}}, \bibinfo {author} {\bibfnamefont {D.~F.~V.}\ \bibnamefont {James}},\ and\ \bibinfo {author} {\bibfnamefont {P.~G.}\ \bibnamefont {Kwiat}},\ }\bibfield  {title} {\bibinfo {title} {Qubit {{Quantum State Tomography}}},\ }in\ \href {https://doi.org/10.1007/978-3-540-44481-7_4} {\emph {\bibinfo {booktitle} {Quantum {{State Estimation}}}}},\ \bibinfo {series} {Lecture {{Notes}} in {{Physics}}}, Vol.\ \bibinfo {volume} {649}\ (\bibinfo  {publisher} {{Springer}},\ \bibinfo {year} {2004})\ \bibinfo {edition} {1st}\ ed.\BibitemShut {Stop}%
\bibitem [{\citenamefont {Poddubny}\ \emph {et~al.}(2016)\citenamefont {Poddubny}, \citenamefont {Iorsh},\ and\ \citenamefont {Sukhorukov}}]{poddubny_generation_2016}%
  \BibitemOpen
  \bibfield  {author} {\bibinfo {author} {\bibfnamefont {A.~N.}\ \bibnamefont {Poddubny}}, \bibinfo {author} {\bibfnamefont {I.~V.}\ \bibnamefont {Iorsh}},\ and\ \bibinfo {author} {\bibfnamefont {A.~A.}\ \bibnamefont {Sukhorukov}},\ }\bibfield  {title} {\bibinfo {title} {Generation of photon-plasmon quantum states in nonlinear hyperbolic metamaterials},\ }\href {https://doi.org/10.1103/PhysRevLett.117.123901} {\bibfield  {journal} {\bibinfo  {journal} {Physical Review Letters}\ }\textbf {\bibinfo {volume} {117}},\ \bibinfo {pages} {123901} (\bibinfo {year} {2016})}\BibitemShut {NoStop}%
\bibitem [{\citenamefont {Santos}\ \emph {et~al.}(2022)\citenamefont {Santos}, \citenamefont {Pertsch}, \citenamefont {Setzpfandt},\ and\ \citenamefont {Saravi}}]{Santos}%
  \BibitemOpen
  \bibfield  {author} {\bibinfo {author} {\bibfnamefont {E.~A.}\ \bibnamefont {Santos}}, \bibinfo {author} {\bibfnamefont {T.}~\bibnamefont {Pertsch}}, \bibinfo {author} {\bibfnamefont {F.}~\bibnamefont {Setzpfandt}},\ and\ \bibinfo {author} {\bibfnamefont {S.}~\bibnamefont {Saravi}},\ }\bibfield  {title} {\bibinfo {title} {Subdiffraction quantum imaging with undetected photons},\ }\href {https://doi.org/10.1103/PhysRevLett.128.173601} {\bibfield  {journal} {\bibinfo  {journal} {Phys. Rev. Lett.}\ }\textbf {\bibinfo {volume} {128}},\ \bibinfo {pages} {173601} (\bibinfo {year} {2022})}\BibitemShut {NoStop}%
\end{thebibliography}%

\end{document}